\documentclass[preprint,preprintnumbers,amsmath,amssymb,12pt,nofootinbib]{revtex4-2}
\usepackage{amsfonts}    % list packages between braces
\usepackage{amssymb}
\usepackage{amsfonts,amssymb}
\usepackage{amsmath}
\usepackage{latexsym}
\usepackage{eepic}
\usepackage{graphicx}
\usepackage[usenames,dvipsnames]{color}
\allowdisplaybreaks

\newcommand{\al}{\alpha}
\newcommand{\pa}{\partial}

\newcommand{\veps}{\varepsilon}

\newcommand{\la}{\lambda}
\newcommand{\ta}{\tau}
\newcommand{\ga}{\gamma}

\newcommand{\om}{\omega}
\newcommand{\de}{\delta}

\newcommand{\De}{\Delta}

\newcommand{\half}{\frac{1}{2}}
\newcommand{\rar}{\rightarrow}
\newcommand{\lrar}{\leftrightarrow}

\newcommand{\non}{\nonumber}

\begin{document}

%\title{Maximal superintegrability and exact-solvability: algebra of integrals and hidden algebra}

\title{Quantum two-dimensional superintegrable systems in flat space:
exact-solvability, hidden algebra, polynomial algebra of integrals}

\author{
Alexander V Turbiner, Juan Carlos Lopez Vieyra\\
Instituto de Ciencias Nucleares, UNAM, \\
 M\'exico, CDMX 04510, Mexico\\[4pt]
turbiner@nucleares.unam.mx\\
vieyra@nucleares.unam.mx\\
%q
%Adrian M Escobar Ruiz\\
%Departamento de F\'isica, UAM,
%M\'exico, CDMX 09340, Mexico\\[4pt]
and \\[8pt]
Pavel Winternitz (deceased)\\
Centre de Recherches Math\'ematiques,\\
Universit\'e de Montreal,\\
Montr\'eal (QC) H3C 3J7, Canada\\
}

\begin{abstract}
In this short review paper the detailed analysis of six two-dimensional quantum {\it superintegrable} 
systems in flat space is presented. 
It includes the Smorodinsky-Winternitz potentials I-II (the Holt potential), the Fokas-Lagerstrom model, 
the 3-body Calogero and Wolfes (equivalently, $G_2$ rational, or $I_6$)  models, and 
the Tremblay-Turbiner-Winternitz (TTW) system with integer index $k$.
It is shown that all of them are exactly-solvable, thus, confirming the Montreal conjecture (2001); 
they admit algebraic forms for the Hamiltonian and both integrals (all three can be written as differential 
operators with polynomial coefficients without a constant term), they have polynomial eigenfunctions 
with the invariants of the discrete symmetry group of invariance taken as variables, they have hidden 
(Lie) algebraic structure $g^{(k)}$ with various $k$, 
and they possess a (finite order) polynomial algebras of integrals. Each model is characterized 
by infinitely-many finite-dimensional invariant subspaces, which form the infinite flag. 
Each subspace coincides with the finite-dimensional representation space of the algebra $g^{(k)}$ 
for a certain $k$. In all presented cases the algebra of integrals is a 4-generated 
$(H, {\cal I}_1, {\cal I}_2, {\cal I}_{12}\equiv[{\cal I}_1, {\cal I}_2])$ infinite-dimensional algebra of ordered monomials of 
degrees 2,3,4,5, which is a subalgebra of the universal enveloping algebra of the hidden algebra.
\end{abstract}

\vskip 2cm
%December 28, 2025
%October 22, 2025
%August 29, 2025
%August 27, 2025
%July 27, 2025

\maketitle

%\pacs{}
\section*{Introduction}

It is well known nowadays that there exist many quantum superintegrable systems, where the number of integrals of motions is larger than the dimension of the configuration space, see \cite{Fris:1965, Wint:1966}, for review \cite{MPW:2013}, with the Hydrogen atom as the most notable example. In 2001 it was formulated a conjecture that all maximally superintegrable quantum systems (in flat space), where the number of integrals of motions is maximal possible, are exactly-solvable \cite{TTW:2001}, while the converse is not true (sometimes, this is called {\it the Montreal conjecture}). To the best of the author's knowledge all attempts to find counter-examples have failed so far. In this review article our main emphasis will be on the two-dimensional (planar) quantum superintegrable systems in flat space with the Hamiltonian $H$ and two algebraically independent integrals $I_1, I_2$. We will focus on an algebra of integrals which turns into a polynomial algebra. 
In total, it will be considered six particular models.

The quantum Hydrogen atom in three-dimensional space with the Hamiltonian of the form
\begin{eqnarray}
\label{Hamiltonian}
 \hat {\cal H} \ = \ \frac{1}{2\,\mu}\,{\hat{\bf  p}^2}\ - \ \frac{\alpha}{r} \ ,
\end{eqnarray}
is a widely known superintegrable system.
Here $\al > 0$ for the case of charges of opposite signs, $\hat{\bf  p} = -i\, \hbar\, \nabla$ is the momentum operator, $r=\sqrt{x^2+y^2+z^2}$; for simplicity, a unit reduced mass, $\mu=1$ and $\hbar=1$ are assumed from now on.

\noindent The angular momentum
\begin{eqnarray}
\label{L}
 \hat{\bf  L} \ = \ {\bf r} \times \hat{\bf  p} \ ,
\end{eqnarray}
and the Laplace-Runge-Lenz vector
\begin{eqnarray}
\label{A}
 \hat{\bf  A} \ = \ \frac{1}{2}(\,\hat{\bf  p} \times \hat{\bf  L}\,-\,\hat{\bf  L} \times \hat{\bf  p}\,) \ - \ \frac{\alpha}{r}\,{\bf r} \ ,
\end{eqnarray}
are integrals of motion: they commute with the Hamiltonian (\ref{Hamiltonian}),
\[
 [\hat{\bf  L} ,\,\hat {\cal H}]\ = \ [\hat{\bf  A} ,\,\hat {\cal H}]\ = \ 0\ .
\]
The vectorial integrals $\hat{\bf  L},\,\hat{\bf  A}$ obey the commutation relations:
\begin{eqnarray}
\label{LA-comm-Rel}
     [\hat{L}_i \,,\,\hat {L}_j] \ = \ i\,\epsilon_{ijk}\,\hat{L}_k \ , \ [\hat{A}_i \,,\,\hat {L}_j] \ = \ i\,\epsilon_{ijk}\,\hat{A}_k \ , \ [\hat{A}_i \,,\,\hat {A}_j] \ = \ -2\,i\,\epsilon_{ijk}\,\hat{L}_k\,\hat {\cal H} \ ,
\end{eqnarray}
where ($i,j,k=x,y,z$). Due to the presence of the quadratic term ($\hat{L}_k\,\hat {\cal H}$) in the rhs 
of the third commutation relation, a infinite-dimensional, quadratic algebra of integrals occurs which 
is generated by seven elements $(\hat {\cal H}, \hat {\bf L}, \hat {\bf A})$, as was discovered 
by Vladimir Fock in the 1930's, see for a discussion \cite{MPW:2013} and also recent paper \cite{TE:2023}. 
This algebra has no $\al$-dependence. Probably, it was the first example of a polynomial algebra of integrals. 
The well-known relations between generating elements 
\begin{eqnarray}
\label{A2}
\hat{\bf  A}^2 \ = \ \al^2 \ + \ 2 \,\hat {\cal H}\,(\,\hat{\bf L}^2 \ + \ 1\,)\ ,
\end{eqnarray}
and
\begin{equation}
\label{AL}
     \hat{\bf  L}\,\cdot\,\hat{\bf A} \ = \ \hat{\bf  A}\,\cdot\,\hat{\bf L} \ = \ 0 \ ,
\end{equation}
appear as the artifacts of the representation of the generating elements 
$(\hat {\cal H}, \hat {\bf L}, \hat {\bf A})$ by differential operators. 
This reflects the fact that in six-dimensional phase space there exist not more 
than five algebraically independent differential operators in three variables.

In \cite{Fris:1965, Wint:1966} based on physics grounds four different two-dimensional 
superintegrable systems were described, 
all of which admitted separation of variables in multiple coordinate systems. A certain list 
of additional superintegrable systems was presented in \cite{Bonatsos,Bonatsos:1993} and \cite{Post:2011}.

This review paper is organized as follows. In Section I the general theory of (maximally) 
superintegrable systems is presented. Section II is about the 4-generated polynomial algebra of integrals 
and its representation theory in differential operators for planar superintegrable systems. The
algebra $g^{(s)}$ is described in Section III. Various 2D superintegrable systems listed 
in Bonatsos et al, \cite{Bonatsos:1993,Bonatsos} are contained in Section IV: 
Smorodinsky-Winternitz Case I (Part A)
and Case II (Part B), Fokas-Lagerstrom (Part C). 
%Higgs I (Part E) and Kepler-Coulomb problem/Higgs II (Part F)
In Section V the celebrated 3-body Calogero model, equivalently, ${A}_2$ rational model 
at $\om=0$ (singular case) and in Section VI the 3-body Wolfes model, equivalently, 
${G}_2/I_6$ rational model at $\om=0$ (singular case). 
Section VII is about the TTW system with integer index $k$. Section VIII is written by AVT and dedicated 
to the memory of Pavel Winternitz, the 3rd coauthor, who recently passed away. 

\newpage

\section{Generalities}

Let us consider a classical mechanical system in $n$-dimensional configuration space
$q \in {\bf R}^n$ with Hamiltonian,
\begin{equation}
\label{Hclas}
  H\ =\ g^{\mu \nu} (q)\, p_{\mu}\, p_{\nu} \ +\ V(q) \ ,
\end{equation}
where the Poisson bracket $\{p_i \,,\, q_j\}\ =\ \de_{i\,j}$ and $\de_{i\,j}$ is the Kronecker symbol, $V(q)$
is the potential, $g^{\mu \nu}$ is the inverse inertia tensor. 
The quantum mechanical counterpart of (\ref{Hclas}) appears as a result of 
the Bruce de Witt factor ordering, see e.g. \cite{Ryan-Tur}, it reads
\[
     g^{\mu \nu} (q)\, p_{\mu}\, p_{\nu}\ \rar\ \frac{1}{\sqrt{g}}\,\pa_{\mu}\,g^{\mu \nu}\,{\sqrt{g}}\,\pa_{\nu}\ ,
\]
and leads to the quantum Hamiltonian
\begin{equation}
\label{Hquant}
  {\cal H}\ =\ -\De_g^{(n)} \ +\ V(q) \ ,
\end{equation}
where $\De_g^{(n)}$ is the Laplace-Beltrami operator on an $n$-dimensional Riemannian or 
pseudo-Riemannian manifold embedded in ${\bf R}^n$, characterized by a contravariant 
metric/cometric $g^{\mu \nu}$,
\begin{equation}
\label{LB}
  \De_{g}\ =\ \frac{1}{\sqrt{g}}\,\pa_{\mu}\,g^{\mu \nu}\,{\sqrt{g}}\,\pa_{\nu}  \ ,
\end{equation}
where $g=\det g_{\mu \nu}$, $\pa_{\mu} = \frac{\pa}{\pa {x_{\mu}}}$ and $V$ is a potential function on the manifold. 
We are mostly focussed on flat space, where the Riemann tensor for $g_{\mu\nu}$ vanishes. For two-dimensional space 
this corresponds to zero curvature. 

We are concerned with the quantum mechanical eigenvalue problems ${\cal H}\Psi=E\Psi$ that 
can be solved exactly, say, using algebraic means, for the eigenvalues and eigenvectors. 
Let us emphasize that the classical and quantum potentials in (\ref{Hclas}) and (\ref{Hquant}), 
respectively, must
coincide (on physics grounds), thus, the potential in (\ref{Hquant}) is $\hbar$ independent. 
Thus, it excludes the so-called $\hbar$-dependent potentials, which are sometimes called also {\it exotic}.

The polynomial in momenta $\{ p \}$ function $I(p,q)$ is called an {\it integral}, if the Poisson bracket vanishes,
\[
   \{ H(p,q) \,,\, I(p,q) \}\ =\ 0\ ,
\]
in the classical case, where $H(p,q) \,,\, I(p,q)$ are functionally independent. 
While in the quantum case a partial differential operator of finite order with variable coefficients 
${\cal I}(p,q)$ is called an {\it integral}, if the Lie bracket (commutator) vanishes,
\[
   [ {\cal H}(p,q) \,,\, {\cal I}(p,q) ]\ =\ 0\ ,
\]
under the assumption that the operators ${\cal H}(p,q) \,,\, {\cal I}(p,q)$ are algebraically independent.

%\pagebreak

{\bf I. (complete)-Integrability}

\begin{itemize}

\item
   We say that a classical system is {\bf (completely)-integrable} if it admits $n$
   functionally independent polynomials in momenta with variable coefficients (integrals)
   $ I_0= H, I_1,\cdots, I_{n-1}$  such that they span the commutative Poisson algebra: $\{I_i,I_j\}=0$
   for $0\le i,j\le n-1$. Here, $\{ A,B\}$ denotes the Poisson bracket of $A, B$.
   The integrals are chosen to be polynomials in $p$ of a minimal order:
   in this case they are called {\bf basic} integrals.

   In this case the phase space splits into (invariant) tori.

\item
   We say that a quantum system is {\bf (completely)-integrable} if it admits $n$
   algebraically independent partial differential operators of finite order with variable coefficients
   ${\cal I}_0={\cal H}, {\cal I}_1,\cdots, {\cal I}_{n-1}$ such that they span the commutative Lie algebra: $[{\cal I}_i,{\cal I}_j]=0$ for $0\le i,j\le n-1$. Here, $[A,B]=AB-BA$ is the commutator (Lie bracket). 
   The integrals are chosen to be the differential operators of a minimal order: in this case they 
   are called {\bf basic} integrals.

   In this case any eigenfunction of the Hamiltonian ${\cal I}_0={\cal H}$ is marked by the quantum 
   numbers = the eigenvalues of the $(n-1)$ commutative integrals ${\cal I}_j,\ j=1,2,\ldots (n-~1)$.
\end{itemize}

{\bf II. (super)-Integrability}

\begin{itemize}

\item
The classical completely-integrable system is {\bf superintegrable} if there are $s \ge 1$ additional 
polynomials in momenta with variable coefficients $I_n,I_{n+1},\cdots,I_{n+s-1}$ such that $\{{H},I_j\}=0$ for $0\le j\le n+s-1$ and the set $\{I_j:\ 0\le j\le n+s-1\}$ is functionally independent. If $s=n-1$, apparently 
the maximum possible, the system is {\bf maximally superintegrable}. If the additional integrals are 
chosen to be polynomials in $p$ of a minimal order: they are called {\bf basic} additional classical 
integrals. In general, additional integrals do not commute with each other or with integrals 
from the commutative algebra $I_i=0$ for $1\le i\le n-1$. Sometimes, the basic integrals $I_j$ 
of a superintegrable system generate a polynomial algebra under the Poisson bracket. 
If the basic integrals are linear in momenta they span a Poisson-Lie algebra.

\item
The quantum completely-integrable system is {\bf superintegrable} if there are $s \ge 1$ additional 
partial differential operators with variable coefficients ${\cal I}_n, {\cal I}_{n+1},\cdots, {\cal I}_{n+s-1}$ such that $[{\cal H}, {\cal I}_j]=0$ for $0\le j\le n+s-1$ and the set 
$\{{\cal I}_j:\ 0\le j\le n+s-1\}$ is algebraically independent. If $s=n-1$, apparently
the maximum possible, the system is {\bf maximally superintegrable}. If the additional 
integrals are chosen to be the differential operators of a minimal order, they are called 
{\bf basic} additional quantum integrals. In general, additional integrals do not commute 
with each other, $[{\cal I}_{n+i}, {\cal I}_{n+j}]\neq 0$ for $i \neq j \geq 0$,  
or with integrals from the commutative algebra ${\cal I}_i$ for $1 \le i \le n-1$, 
$[{\cal I}_{i}, {\cal I}_{n+j}] \neq 0$ for $j \geq 0$. In particular, for two-dimensional 
case $[{\cal I}_{1}, {\cal I}_{2}] \equiv {\cal I}_{12} \neq 0$.
The basic integrals  ${\cal I}_j$ of a superintegrable system can generate a polynomial 
algebra under the Lie bracket/commutator, which is not usually a Lie algebra. However, 
if all basic integrals are first order differential operators, they span a Lie algebra. 
The simplest example of this situation is the isotropic harmonic oscillator.

\end{itemize}

{\bf III. (Quantum) Solvability}

There are two principal types of quantum systems that can be solved exactly:

\begin{itemize}

\item
   A quantum system is {\bf exactly-solvable, (ES)} if there is an infinite flag of subspaces 
   ${\cal P}_{j}$, $j=1,2, \cdots$, of the domain of ${\cal H}$
   such that $n_j=\dim {\cal P}_{j}\to \infty$ as $j \to \infty$ and
   ${\cal H}{\cal P}_{j}\subseteq {\cal P}_{j}\subset {\cal P}_{j+1}$ for each $j$. Note that 
   for each subspace ${\cal P}_{j}$ the $n_j$ eigenvalues and eigenfunctions of ${\cal H}$ 
   can be determined by algebraic means.

\item
   A quantum system is {\bf quasi-exactly solvable, (QES)} if there is a single  subspace
   ${\cal P}_k$ of dimension $n_k>0$ such that
   ${\cal H} {\cal P}_{k}\subseteq {\cal P}_{k}$. In this case we can find  $n_k$ eigenvalues 
   and eigenfunctions of ${\cal H}$ by algebraic means, but we have
   no information obtained by algebraic means about the remaining eigenvalues and eigenfunctions.

\end{itemize}

\bigskip

{\bf CONJECTURE}

\bigskip

(A) For any classical {\it maximally superintegrable} system all bounded trajectories 
are closed and periodic.

(B) Any quantum {\it maximally superintegrable} system on the Eucledian plane 
is exactly-solvable.

\newpage

{\bf COMMENTS}

\bigskip

\begin{itemize}

\item
 The structure and representation theory of the {\bf algebra of integrals} provides information about the spectral decomposition of the quantum system. In particular, maximal superintegrability captures the properties of quantum Hamiltonian systems that allow the Schr\"odinger eigenvalue problem $H\Psi=E\Psi$ to be solved exactly, analytically and algebraically.

\item
   Let ${\mathbf U_h}$ be an algebra of differential operators  that is  finitely-generated by ${\mathbf h}$. If ${\mathbf h}$ is a Lie algebra, ${\mathbf U_h}$ is its universal enveloping algebra. We say that a quantum system has {\bf hidden} algebra ${\mathbf U_h}$ if the Hamiltonian $I_0={\cal H}$ is an element of ${\mathbf U_h}$. In all so far known examples of this,  not only the Hamiltonian but all integrals are elements of ${\mathbf U_h}$. In this case the algebra of integrals is a sub-algebra of the hidden algebra. A trivial example of a hidden algebra is the Heisenberg-Weyl algebra. The first non-trivial example is  ${\mathbf h} =sl_2$ realized by the first order differential operators on $RP^1$. It is the explanation for the (quasi)-exact-solvability of many one-dimensional Schr\"odinger operators. In general, if the hidden algebra ${\mathbf U_h}$ has a finite-dimensional representation, the Hamiltonian (and sometimes integrals) has a finite-dimensional invariant subspace which coincides with finite-dimensional representation space of the hidden algebra. The Schr\"odinger eigenvalue problem ${\cal H}\Psi=E\Psi$ can be solved by algebraic means for the elements of
   the finite-dimensional representation space.

\end{itemize}

Quantum systems and their classical analogs that can be solved exactly have been  
of enormous historical importance: the harmonic oscillator,
the Kepler system (and the Hohmann transfer, used in celestial navigation), the quantum 2-body Coulomb system and, in particular, the hydrogen atom
(and its use to develop a perturbation theory for the periodic table of the elements), etc. The discovery and analysis 
of such systems is clearly of importance, see for a historical account and discussion a remarkable review paper \cite{MPW:2013}. 
For example, in \cite{TTW:2001} there is a conjecture that all 2nd order superintegrable systems in Euclidean space 
$E^n$ are exactly solvable.

\section{Polynomial algebras of (integrals).}

We call {\it the polynomial algebra of integrals} the infinite-dimensional, 4-generated, associative 
algebra of ordered monomials
\begin{equation}
\label{PAI}
  H^n\,{\cal I}_1^m\,{\cal I}_2^p\, {\cal I}_{12}^q\ ,
\end{equation}
where $n, m, p, q$ are non-negative integers, with the generating elements 
$(H, {\cal I}_1, {\cal I}_2, {\cal I}_{12})$ obeying the following commutation/structure relations:  
\[
 [H,{\cal I}_1] = [H,{\cal I}_2] = 0\ ,
\]
with
\[
{\cal I}_2\,{\cal I}_1\ =\ - {\cal I}_{12} + 2 {\cal I}_1\,{\cal I}_2\ ,
%  {\cal I}_{12} \equiv [{\cal I}_1,{\cal I}_2] \ ,
\]
and
\[
  [H,{\cal I}_{12}]\ =\ 0\ .
\]
The ``double" commutators $[{\cal I}_1,{\cal I}_{12}]$ and  $[{\cal I}_2, {\cal I}_{12}]$ 
\begin{equation}
\label{I1-I12abst}
  [{\cal I}_1,{\cal I}_{12}]\ =\ P (H, {\cal I}_1,{\cal I}_2,{\cal I}_{12}) \ ,
\end{equation}
\begin{equation}
\label{I2-I12abst}
  [{\cal I}_2,{\cal I}_{12}]\ =\ Q (H, {\cal I}_1,{\cal I}_2,{\cal I}_{12}) \ ,
\end{equation}
are given by finite-degree polynomials in $H, {\cal I}_1,{\cal I}_2,{\cal I}_{12}$ each of them is written 
as a combination of ordered monomials \footnote{We choose a lexicographical ordering: 
$H, {\cal I}_1,{\cal I}_2,{\cal I}_{12}$.}. 
Degree of the algebra is defined by maximal degree of $P,Q$ polynomials.
This algebra in such a form was defined in \cite{Turbiner:2024}.

By taking a planar, two-dimensional superintegrable system (in the flat space) one can construct 
the representation theory of (\ref{PAI}) by realizing the generating elements 
$(H, {\cal I}_1, {\cal I}_2, {\cal I}_{12})$ as differential operators by
identifying them as the Hamiltonian, two integrals and the commutator of integrals 
${\cal I}_{12}=[{\cal I}_{1}, {\cal I}_{2}]$.
Such a realization of the polynomial algebra of integrals was proposed in \cite{Miller:2010}.
By calculating the double commutators $[{\cal I}_1,{\cal I}_{12}]$ (\ref{I1-I12abst}) and  
$[{\cal I}_2, {\cal I}_{12}]$ (\ref{I2-I12abst}) one
can find the finite-degree polynomials in $H, {\cal I}_1,{\cal I}_2,{\cal I}_{12}$,
\begin{equation}
\label{I1-I12}
  [{\cal I}_1,{\cal I}_{12}]\ =\ P_n (H, {\cal I}_1,{\cal I}_2,{\cal I}_{12}) \ ,
\end{equation}
\begin{equation}
\label{I2-I12}
  [{\cal I}_2,{\cal I}_{12}]\ =\ Q_m (H, {\cal I}_1,{\cal I}_2,{\cal I}_{12}) \ ,
\end{equation}
which specify the concrete polynomial algebra of integrals. It turns out that for a number of 
studied planar superintegrable systems the polynomials $P,Q$ are of the same degree, 
$m=n$, while for some other cases one of the polynomials is of the first degree, $n=1$. 
It has to be mentioned that in the case of physics systems the Hamiltonian $H$ and 
the integral ${\cal I}_1$, responsible for separation of variable, is of the second order,
are defined uniquely on physics grounds.
The second integral ${\cal I}_2$ admits a certain ambiguity: 
(i) in its basic form it should be of the minimal degree differential operator and 
(ii) a non-linear combination of $H, {\cal I}_1$, which does not exceed the minimal degree 
of ${\cal I}_2$, can be added. There exists ${\cal I}_2$ of the minimal degree for which 
the $P,Q$ polynomials are of minimal degrees.
If ${\cal I}_2$ is of non-minimal degree, resulting polynomial algebra of integrals (\ref{PAI}), 
realized by differential operators, appears incomplete. We assume that in this case 
the $P,Q$ polynomials are not of the minimal degree, at least, see \cite{Miller:2011} 
and a discussion in Conclusions.
It must be also emphasized that in four-dimensional phase space there exist at most 
three algebraically independent operators (or integrals of motion), 
where the Hamiltonian is included. 
It implies the existence of the {\it syzygy}: differential operators 
$(H, {\cal I}_1, {\cal I}_2, {\cal I}_{12})$ must be algebraically/polynomially related!

In the physics problems the configuration space is usually parametrized by (natural) Cartesian 
coordinates $(x,y)$ or polar coordinates $(\rho, \varphi)$. It leads to the Hamiltonian 
and integrals defined in those coordinates. However, if the system is invariant wrt a discrete 
symmetry group (reflections, permutation, dihedral $I_{2k}$) natural coordinates are invariants 
of the symmetry group. In this case the gauge rotated Hamiltonian and integrals appear 
in the form of algebraic operators = differential operators with polynomial coefficients. 
Such resulting discrete-symmetry-invariant Hamiltonian and integrals generate a polynomial algebra 
of integrals, which is a sub-algebra of the original algebra of integrals.

\section{$g^{(s)}$ algebra}
\label{gk}

Around 1880 Sophus Lie studied the (Lie) algebras acting on the Euclidian plane $E_2$ 
realized as first order differential operators in two variables. He discovered that the non-semi-simple Lie algebra $g\ell(2,{\bf R}) \ltimes {\cal R}^{(s)}$ 
with integer index $s=1,2,3,\ldots$ acts on $E_2$ as the algebra of vector fields, 
see Case 24 in \cite{Lie:1880}.
In \cite{gko:1992,gko:1994} the Lie results were extended to the $g\ell(2,{\bf R})$ algebra of  first order differential operators,
\[
J^1\  =\  \pa_t \ ,
\]
\begin{equation}
\label{gl2r}
  J^2_N\  =\ t \pa_t\ -\ \frac{N}{3} \ ,
 \ J^3_N\  =\ s u\pa_u\ -\ \frac{N}{3}\ ,
\end{equation}
\[
    J^4_N\  =\ t^2 \pa_t \  +\ s t u \pa_u \ - \ N t\ ,
\]
where $N$ is a real number (the case of vector fields corresponds to $N=0$), and
\begin{equation}
\label{R}
   R_{i}\  = \ t^{i}\pa_u\ ,\ i=0,1,\dots, s\ ,\quad {\cal R}^{(s)}\equiv (R_{0},\ldots, R_{s})\ ,
\end{equation}
which spans the commutative algebra ${\cal R}^{(s)}$. For $s=1$ this algebra becomes 
the subalgebra of the algebra $s\ell(3)$. If $N$ is a non-negative integer, there exists a finite-dimensional representation of (\ref{gl2r}), (\ref{R}) i.e. 
\begin{equation}
\label{space_r}
 {\cal P}_{{\cal N}}^{(s)}\ =\ (t^{p} u^{q} | 0\leq (p + s q) \leq {\cal N})\ , \quad {\cal N} = 0,1,2,\ldots  \ ,
\end{equation}
where the algebra acts reducibly. Those finite-dimensional subspaces 
${\cal P}_{{\cal N}}^{(s)}$ can be ordered by forming an {\it infinite flag},
\begin{equation}
\label{flag}
  {\cal P}_0^{(s)} \subset {\cal P}_1^{(s)} \subset {\cal P}_2^{(s)} \ldots {\cal P}_{\cal N}^{(s)} 
  \ldots\ \equiv\ {\cal P}^{(s)}\ .
\end{equation}

In \cite{Turbiner:1998,TTW:2009} it was shown that adding to (\ref{gl2r}), (\ref{R}) a single higher order differential operator,
\begin{equation}
\label{grT}
 T^{(s)}_0\ =\ u\pa_{t}^s\ ,
\end{equation}
makes the action on ${\cal P}_{{\cal N}}^{(s)}$ irreducible. Furthermore, 
by taking multiple commutators 
\begin{equation}
\label{grT-subalgebra}
    T_i^{(s)}\ =\ \Bigr[ \underbrace{J^4,\bigr[J^4,[ \ldots , [J^4,T_0^{(s)}]]\bigr]}_i \Bigr] =
    u\pa_{t}^{s-i} J_0 (J_0+1)\ldots (J_0+i-1) \ ,
\end{equation}
with $i=1,\ldots s$,\ we generate the differential operators of the degree $s$ acting 
on the space ${\cal P}_{{\cal N}}^{(s)}$ (\ref{space_r}), see \cite{Turbiner:2010} and 
also \cite{ST:2013}. 
Here 
\begin{equation}
\label{j0}
 J_0=t\pa_t \  +\ s u\pa_u \ - \ N\ ,
\end{equation} 
is the Euler-Cartan generator (or, in physics language, the number operator), which maps a monomial 
in $(t,u)$ to itself. Interestingly, there is a property of nilpotency,
\[
  T_i^{(s)} = 0\ ,\ i > s \ .
\]
The non-trivial $T$-operators span the commutative algebra,
\[
  [T_i^{(s)},T_j^{(s)}] = 0\ ,\quad i,j=0,\ldots s\ ,\quad {\cal T}^{(s)}\equiv (T_{0}^{(s)},\ldots, T_{s}^{(s)})\ ,
\]
of dimension $(s+1)$. Let us generate its structure in the form of the Gauss decomposition:
\[ 
{\cal T}^{(s)} \rtimes gl_2 \ltimes {\cal R}^{(s)}
\]
which corresponds to the following diagram
\begin{center}
%\fbox{
\begin{picture}(100,50)
%%%%%%%%%%%%%%%%%%%%%%%%%%%%%%%%
\put(50,0){\vector(1,0) {18}}
\put(50,0){\vector(-1,0){18}}
\put(50,0){\vector(0,1) {15}}
%%%%%%%%%%%%%%%%%%%%%%%%%%%%%%%%
\put(48,32){$g\ell_2$}
\put(30,28){\rotatebox{225}{\Large $\ltimes$}}
\put(68,23){\rotatebox{-45}{\Large $\ltimes$}}
\put(7,0){$\scriptstyle {\cal R}^{(s)}$}
\put(87,0){$\scriptstyle {\cal T}^{(s)}$}
\put(35,-17){${ P}_s{ (g\ell_2)}$}
%%%%%%%%%%%%%%%%%%%%%%%%%%%%%%%%%
\end{picture}
%}
\end{center}
where ${P}_s{ (g\ell_2)}$ is a polynomial of degree $s$ in the $g\ell_2$ generators.
The dimension of the structure is $(2s+6)$. 
For $s=1$ it is the true Gauss decomposition of the $s\ell(3)$ algebra. By definition {\it the $g^{(s)}$ algebra is the infinite-dimensional,\ $(2s+6)$-generated associative algebra of differential operators with ${\cal P}_n^{(s)}$ as its finite-dimensional irreducible representation space}. In particular, $g^{(1)} = U_{gl(3)}$, thus, it coincides with the universal enveloping algebra of $gl(3)$-algebra.

\section{Various two-dimensional superintegrable systems (from Bonatsos et al (five systems))}

%{\color{red} \bf ADRIAN  }

\subsection{Smorodinsky-Winternitz system, Case I (1965)}

Hamiltonian \cite{Fris:1965,Wint:1966} in Cartesian coordinates has the form
\begin{equation}
\label{CaseI}
  H \ = \ -\frac{1}{2}(\,\pa_x^2 \ + \  \pa_y^2\,) \ + \ \frac{\om^2}{2}\,(x^2 \ + \ y^2) \ + \ \frac{A}{x^2}\ + \ \frac{B}{y^2} \ ,
\end{equation}
where $\om > 0$ and $A,B > -\frac{1}{8}$ are parameters, $\pa_x \equiv \frac{\pa}{\pa x}, \pa_y \equiv \frac{\pa}{\pa y}$. 
It is known in literature as the Smorodinsky-Winternitz system I (SW-I). This system admits two integrals
\begin{equation}
\label{ACaseI}
          {\cal I}_1 \ = \ -\pa_y^2 \ + \ \om^2\,y^2 \ + \ \frac{2\,B}{y^2}\ ,
\end{equation}
and
\begin{equation}
\label{BCaseI}
{\cal I}_2 \ = \ -\,x^2\,\pa_y^2 \ - \  y^2\,\pa_x^2  \ + \ 2\,x\,y\,\pa_{x,y}^2 \ + \ x\,\pa_x  \ + \ y\,\pa_y \ + \ \frac{2\,B\,x^2}{y^2} \ + \ \frac{2\,A\,y^2}{x^2} \ + \ 1     \ ,
\end{equation}
see e.g. \cite{Bonatsos}. It can be immediately verified that $[H,{\cal I}_1]=[H,{\cal I}_2]=0$. It is worth noting 
that the SW-I system admits the separation of variables in polar coordinates, thus, the integrals can be written 
differently, see Section VII (about TTW), the case $k=1$.

The discrete symmetry of the Hamiltonian (\ref{CaseI}) and integrals (\ref{ACaseI})-(\ref{BCaseI}) in Cartesian 
coordinates is reflections ${Z_2}^{\oplus 2}$ ($x \rightarrow -x,\,y \rightarrow -y $). 
It implies that the parities $p_x$ and $p_y$ wrt $x$ and $y$, respectively, of the eigenstates can be introduced. The eigenfunctions have the form,
\[
    \Psi\ =\ x^{p_x} y^{p_y} \psi(x^2, y^2)\quad ,\ p_x, p_y = 0,1 \ .
\]
Hence, there exist four families of states: even-even ($p_x=p_y=0$), even-odd ($p_x=0, p_y=1$), odd-even ($p_x=1, p_y=0$) and
odd-odd ($p_x=1, p_y=1$).  
Note that in the case $A=B$ the symmetry is extended to ${Z_2}^{\oplus 2}\otimes S_2$ (reflections plus
permutation $x \lrar y$).

Let us introduce the commutator of integrals (\ref{ACaseI}), (\ref{BCaseI})
\begin{equation}
\label{AB-CaseI}
{\cal I}_{12} \ \equiv \  [{\cal I}_1,\,{\cal I}_2] \ ,
\end{equation}
which is a differential operator of third order, see e.g. \cite{TTW:2001},
\begin{eqnarray}
&{\cal I}_{12} \ = \  4\,y\,\pa^2_x\,\pa_y \ - \ 4\,x\,\pa^2_y\,\pa_x \ + \ 2\,\pa_x^2 \ - \ 2\,\pa_y^2
\\ &
- \ 4\,x\,\bigg(\om^2 y^2\,-\,\frac{2\, B}{y^2}\bigg)\,\pa_x \ + \ 4\,y\,\bigg(\om^2 x^2\,-\,\frac{2 \,A }{x^2}\bigg)\,\pa_y \ - \ \frac{4\, A}{x^2}+\frac{4\, B}{y^2}+2\, \om^2 (x^2-y^2) \non
\ .
\end{eqnarray}
It evidently commutes with Hamiltonian (\ref{CaseI}), $[H, {\cal I}_{12}]=0$\,.

Double commutators are easily calculated
\begin{equation}
\label{double-1-CaseI}
[{\cal I}_{1},\,{\cal I}_{12}] \ = 8\,{\cal I}_1^2 \ - \ 16\,H\,{\cal I}_1 \ + \ 16\,\om^2\,{\cal I}_2 \ - \ 8\,\om^2 \ ,
\end{equation}
\begin{equation}
\label{double-2-CaseI}
 [{\cal I}_{2},\,{\cal I}_{12}] \ = \   -16\,{\cal I}_1\,{\cal I}_2 \ + \ 16\,H\,{\cal I}_2 \ - \ 8\,{\cal I}_{12} \ - 
 \ 32\,(A+B-1)\,{\cal I}_1 -32\,(1-2\,B)\,H  \ ,
\end{equation}
they are second order polynomials in $H,\,{\cal I}_1,\,{\cal I}_2,\,{\cal I}_{12}$. The rhs of (\ref{double-1-CaseI}) does not depend on ${\cal I}_{12}$. Hence, we arrive at a quadratic algebra of integrals 
generated by ($H,\,{\cal I}_1,\,{\cal I}_2,\,{\cal I}_{12}$) realized by differential operators. Interestingly, the algebra remains quadratic 
even if $\om=A=B=0$, which corresponds to the flat Laplacian: the potential in (\ref{CaseI}) vanishes.  

Since in four-dimensional phase space there exist at most
three algebraically independent operators (or, stating differently, integrals of motion), where the Hamiltonian
is included, it implies the existence of a syzygy: four differential operators ($H, {\cal I}_1, {\cal I}_2, {\cal I}_{12}$) 
are algebraically related. In our particular case the syzygy is realized via a polynomial relation,
\begin{equation*}
 {\cal I}_{12}^2 = {\cal R}(H, {\cal I}_1, {\cal I}_2, {\cal I}_{12})\ ,
\end{equation*}
where ${\cal R}$ has the form of the cubic polynomial,
\begin{align}
 {\cal I}_{12}^2 & =
 16 {\cal I}_1^2{\cal I}_2\ -\ 32 H {\cal I}_1 {\cal I}_2\ +\ 16 H {\cal I}_{12}\ -\ 16 {\cal I}_1 {\cal I}_{12}\ 
 +\ 16(8 B - 3) H^2\ -\ 64 (2 B-1) H {\cal I}_1 \non
 \\ &
 +\ 32 (A + B - 1) {\cal I}_1^2\ +\ 16\om^2 {\cal I}_2^2\ +\ 112\om^2 {\cal I}_2\ -\ 32\om^2 (8 A B - 3 A - 3 B + 3) \ ,
\end{align}
with explicit dependence on ${\cal I}_{12}$.
\bigskip

Let us introduce a gauge factor
\begin{equation}
\label{Gamma-I}
   \Gamma \ = \ e^{-\frac{\om}{2}(\ta_1+\ta_2)}\, \tau _1^{\frac{1}{4}+\frac{1}{4}k_1\, \sqrt{1+8\, A }}\,\tau _2^{\frac{1}{4}+\frac{1}{4}k_2\, \sqrt{1+8\, B }} \quad , \qquad k_{1,2}=\pm 1 \ ,
\end{equation}
which has a meaning of the ground state function of (\ref{CaseI}), 
where
\[
\ta_1 \ = \ x^2\quad , \quad \ta_2 \ = \ y^2 \ ,
\]
are new variables: they are invariants of ${Z_2}^{\oplus 2}$ group, which realizes the symmetry of both the Hamiltonian (\ref{CaseI}) and integrals.
It leads to gauge-rotated Hamiltonian and integrals in the form of differential operators with polynomial coefficients 
\footnote{Since now on we will call them the algebraic operators.}:
\begin{eqnarray}
\label{h-1}
& h \  \equiv \ \Gamma^{-1}\,(H \,- \,E_{0})\,\Gamma
\\ & \ = \ -2\,\tau_1\,\pa^2_{\tau_1}\ - \ 2\,\tau_2\,\pa^2_{\tau_2} \ - \ (\beta_1 \,-\,2\,\tau_1\,\om\,)\,\pa_{\tau_1} \ - \ (\beta_2 \,-\,2\,\tau_2\,\om\,)\,\pa_{\tau_2}  \ ,\non
\end{eqnarray}
here $E_{0}  =  \frac{\om\,(\beta_1+\beta_2)}{2} $ is the ground state energy, $\beta_1 \equiv (2+k_1\,\sqrt{1+8\,A })$ and 
$\beta_2 \equiv (2+k_2\,\sqrt{1+8\,B })$, it is sum of two Laguerre operators, and
\begin{equation}
\label{i1-1}
  i_1 \  \equiv \ \Gamma^{-1}\, ({\cal I}_1\,-\,{\cal I}_1^{(0)})\, \Gamma
 \ = \  -4\,\tau_2\,\pa^2_{\tau_2} \ - \ 2\,(\beta_2\,-\,2\,\tau_2\,\om)\,\pa_{\tau_2} \ ,
\end{equation}
where ${\cal I}_1^{(0)}=(\beta_2\,\om)$ is the lowest eigenvalue of ${\cal I}_1$, it is a Laguerre operator, and 
\begin{eqnarray}
\label{i2-1}
& i_2 \  \equiv \ \Gamma^{-1}\,\bigg({\cal I}_2 - {\cal I}_2^{(0)} \bigg)
\,\Gamma
\\ &
\ = -4\,\tau_1\,\ta_2\,( \pa^2_{\ta_1}\,+\, \pa^2_{\ta_2}\,-\, 2\,\pa_{\ta_1}\,\pa_{\ta_2} ) \ - 
\ 2 \left(\beta_2 \ta_1 -\beta_1 \ta_2 - 4 \ta_1 + 4 \ta_2 \right)\,(\pa_{\ta_1}\,-\,\pa_{\ta_2})    \
\ ,\non
\end{eqnarray}
where ${\cal I}_2^{(0)}=\frac{1}{2}(\beta_1\,\beta_2 -4\beta_1-4\beta_2+17)$ is the lowest eigenvalue of ${\cal I}_2$, and 
\begin{eqnarray}
\label{i12-1}
& i_{12} \  \equiv \ \Gamma^{-1}\,{\cal I}_{12}\,\Gamma
\\ & \ = \  32\,\tau_1\,\tau_2\,(\pa^2_{\tau_1}\pa_{\tau_2} \,-\,\pa^2_{\tau_2}\pa_{\tau_1})
 \ -\ 8\, \ta_1\, \left(\beta_2-4+2\, \ta_2 \,\om \right)\,\pa^2_{\ta_1} \ + 
 \ 8 \ta_2 \left(\beta _1+2 \ta_1 \om -4\right)\,\pa^2_{\tau_2} \non
\\ &
+ \ 16 \left(\beta_2 \ta_1-\beta_1 \ta_2-4 \ta_1+4 \ta_2\right)\pa^2_{\ta_1,\ta_2} \ +
 \ 4 \left(\beta_1-4\right) \left(\beta _2+2 \tau _2 \om -4\right)\,\pa_{\tau_1} \non
\\ &
- \ 4 \left(\beta _2-4\right) \left(\beta _1+2 \tau _1 \om - 4\right)\,\pa_{\ta_2}
 \ , \non
\end{eqnarray}
see \cite{TTW:2001}. All four operators (\ref{h-1})-(\ref{i12-1}) have no constant terms (the terms containing no derivatives). Their lowest eigenfunctions are constants and the lowest eigenvalues are zeroes. 
Evidently, $i_1, i_2, i_{12}$ commute with the algebraic Hamiltonian $h$, their double 
commutators are at most of the 2nd degree polynomials in $h, i_1, i_2, i_{12}$ but without 
constant terms, cf. (\ref{double-1-CaseI})-(\ref{double-2-CaseI}). All four algebraic operators 
$h, i_1, i_2, i_{12}$ are elements of the universal enveloping algebra of the 5-dimensional 
Heisenberg algebra in the coordinate-momentum representation $(\pa_{\ta}, \ta)$.   

In terms of the maximal affine subalgebra $b_3 \in sl(3,R)$ generators
\begin{equation}
\label{b3}
  {\cal J}_i^{-} \ = \ {\pa_{\ta_i}} \quad , \quad  {\cal J}_{ij}^{0} \ = \ 
  \ta_i\,{\pa_{\ta_j}} \quad , \quad i=1,2 \ ,
\end{equation}
we arrive at the Lie algebraic form of the Hamiltonian and integrals:
\begin{eqnarray}
 h \ =  \ -2\,{\cal J}_{11}^{0}\,{\cal J}_1^{-}\ - \ 2\,{\cal J}_{22}^{0}\,{\cal J}_2^{-} \ - \ \beta_1\,{\cal J}_1^{-}  \ - \ \beta_2\,{\cal J}_2^{-} \ + \ 2\,\om\,({\cal J}_{11}^{0}\,+\,{\cal J}_{22}^{0}) \ ,\non
\end{eqnarray}
\begin{eqnarray}
 & i_1 \ =  \  - \ 4\,{\cal J}_{22}^{0}\,{\cal J}_2^{-} \ - \ 2\,\beta_2\,{\cal J}_2^{-}\ + \
4\,\om\,{\cal J}_{22}^{0}
  \ ,\non
\end{eqnarray}
\begin{eqnarray}
 & i_2 \ = \  -4(\,{\cal J}_{11}^{0}\,{\cal J}_{21}^{0}\,+\,{\cal J}_{22}^{0}\,{\cal J}_{12}^{0}\, -\,2\,{\cal J}_{11}^{0}\,{\cal J}_{22}^{0} ) \ - \ 2\,(\beta_2-4)\,{\cal J}_{11}^{0}\ - \ 2\,(\beta_1-4)\,{\cal J}_{22}^{0}  \ ,\non
\end{eqnarray}
as bilinear combinations of generators, see \cite{TTW:2001}, while $i_{12}$ is trilinear. 
It implies that the SW-I system has the hidden algebra $sl(3,R) \in g^{(1)}$. 

Spectra of $h$:
\begin{equation}
 \veps_{n_1,n_2} \ = \ 2\,\om\,(\,n_1\,+\,n_2\,) \ ,
\end{equation}
is characterized by the quantum numbers $n_1,n_2=0,\,1,\,2,\ldots$, it corresponds to the isotropic harmonic oscillator 
with polynomial eigenfunctions given by,
\begin{equation}
    P_{n_1,n_2}(\tau_1,\,\tau_2) \ = \ L_{n_1}^{\small(\frac{1}{2}k_1\,\sqrt{1+8\,A})}(\om\,\tau_1)\,
    L_{n_2}^{\small(\frac{1}{2}k_2\,\sqrt{1+8\,B})}(\om\,\tau_2) \ ,
\end{equation}
which is the product of two generalized Laguerre polynomials.

\subsubsection*{Other considerations}

The algebra of integrals for superintegrable Smorodinsky-Winternitz system, Case I 
was also studied in \cite{Miller:2005,Post:2011}, 
where the Hamiltonian appeared in slightly different forms: for instance, in \cite{Miller:2005}, 
the parameters of the Hamiltonian (\ref{CaseI}) 
are written as: $A=\frac{1}{2}(k_1^2 - \frac{1}{4})$, and $B=\frac{1}{2}(k_2^2 -\frac{1}{4})$, 
where $k_1,k_2$ are arbitrary parameters. 
In \cite{Post:2011}, the Hamiltonian appears with overall negative sign, multiplied by a factor 2, 
{\it i.e.} $H^{\mbox{\tiny\cite{Post:2011}}} = -2H$, (where $H$ is given by eq.(\ref{CaseI}) 
with $A=\frac{1}{2}(a^2-\frac{1}{4})$ and $B=\frac{1}{2}(b^2-\frac{1}{4})$, with $a,b$ as parameters). 
In the following we will adopt the form of the Hamiltonian (\ref{CaseI}) and we will make appropriate 
changes in the formulas presented in \cite{Miller:2005,Post:2011} to which we will address below.

In \cite{Miller:2005}, three integrals (commuting with the Hamiltonian $H$ (\ref{CaseI})) are
\begin{equation}
  L_1 = \pa_x^2 -  \frac{2 A}{x^2} -\om^2\, x^2\ ,\quad
 L_2 = \pa_y^2 -  \frac{2 B}{y^2} -\om^2\, y^2\ \equiv -{\cal I}_1 ,
\end{equation}
\begin{equation}
  L_3 = (x\pa_y - y\pa_x)^2 - 2 A \frac{y^2}{x^2} - 2 B \frac{x^2}{y^2} - \frac{1}{2}\ \equiv - {\cal I}_2 + 1/2,
\end{equation}
cf.(\ref{ACaseI})-(\ref{BCaseI}), and the commutator
\[
-{\cal I}_{12} \equiv R \, = \, [L_1,L_3]\, =\,  -[L_2,L_3]\ ,
\] 
cf.(\ref{AB-CaseI}). Notice that $L_1$, $L_2$ are not independent integrals due to the relation $-2H=L_1+L_2$ 
(see \cite{Miller:2005}).
The algebra of the integrals is defined as \footnote{In \cite{Miller:2005} the first two double commutators are  
incorrectly described as $[L_i,R]  = -4\{L_i,L_j\} + 16\om^2 L_3$,  \hbox{$ i\neq j, \ i,j=1,2 $}.}
\begin{align*}
 [L_1,R] & = \phantom{-}4 \{L_1,L_2\} + 16\om^2 L_3  
\\& 
 = -16 H L_1 - 8 L_1^2 + 16\om^2 L_3\ ,
\\  
 [L_2,R] & =  
 -4 \{L_1,L_2\} - 16\om^2 L_3 
\\& 
 = \phantom{-}16 H L_1  + 8 L_1^2 - 16\om^2 L_3 \,,
\\ 
 [L_3,R] & = \phantom{-}4\{L_1,L_3\} - 4\{L_2,L_3\} + 2(3-8 B)L_1 - 2(3-8 A)L_2 \\
         & =\ 16 H L_3 + 16 L_1 L_3 + 8 (3-8 A) H + 8(3-4 (A+B))L_1 - 8 R \ , 
\end{align*}
where, in the second line of each commutator the r.h.s. is written as a polynomial 
of ordered monomials in $(H,L_1,L_3,R)$.
The syzygy relation is
\begin{align*} 
 R^2 & = (8/3)\{L_1,L_2,L_3\} + (64/3) \{L_1,L_2\} + 16\om^2 L_3^2 - 4 (3 - 8 B) L_1^2 - 4 (3 - 8 A) L_2^2  
\\&
 \ -\ (128/3) \om^2 L_3 - 4 \om^2 (3 - 8 A)(3 - 8 B)
\\& 
 = -32 H L_1 L_3 - 16 L_1^2 L_3 + 16 H R + 16 L_1 R - 16 (3 - 8 A)\,H^2 - 16 (3 - 8 A)\,H L_1 
\\&
 + 8 (4(A+B)-3)\,L_1^2 + 16\om^2 L_3^2 - 128\om^2 L_3 - 4\om^2 (3-8 A)(3-8 B) \ ,
\end{align*}
where $\{,\}$ denotes anticommutator and $\{, ,\}$ is the so-called triple symmetrizer 
\footnote{For definitions, see \cite{Miller:2005}.  
Explicitly, $\{A,B\} = -[A,B] + 2 AB$, triple symmetrizer $\{ A,B,C \equiv [A,B] \}$ can be also written 
as a combination of ordered monomials in $(A, B, [A,B])$, however, after reordering it leads to quite 
complicated expression.}. The algebra of integrals is quadratic.
 
In \cite{Post:2011}, two integrals (commuting with the Hamiltonian $H$ (\ref{CaseI})) are written as
\[
 L_1=\pa_x^2-\frac{2 A}{x^2} - \om^2 x^2\quad ,\quad L_2=(x\pa_y-y\pa_x)^2 - 
 2 A \frac{y^2}{x^2} - 2 B\frac{x^2}{y^2}\ .
\]
The double commutators are \footnote{In \cite{Post:2011} there are two misprints in the signs of the last two terms 
in the rhs of $[R,L_1]$. Here, they are corrected.}, 
\begin{align*}
 \left[R,L_1\right] &= 8L_1^2+16 H L_1 - 16\om^2L_2 + 8\om^2\ ,
\\
 \left[R,L_2\right] &=
 - 16 H L_2 - 8\lbrace L_1, L_2 \rbrace
 + 16\big(4 A - 1\big) H + 16\big(2A+2B-1\big)L_1\ ,
\\
 R^2 & =-16H\lbrace L_1,L_2\rbrace-\frac{8}{3}
 \lbrace L_1,L_1, L_2\rbrace
  + 16\om^2 L_2^2 + 16\big(8A-3\big) H^2
  + 16\left(2(A+B)-\frac{11}{3} \right)L_1^2\non
\\ &
  + 32\left(4 A-\frac{11}{3} \right)HL_1
  - \frac{176\om^2}{3}L_2 - \frac{32\om^2}{3}\big(14 A B - 9 A - 9 B + 1\big)\ ,
\end{align*}
Here $R \equiv \left[L_1, L_2 \right]$. Non-surprisingly, the algebra of integrals remains quadratic.

Concluding this section we have to add that in papers \cite{Fris:1965,Wint:1966} two more superintegrable systems 
were introduced, called in literature as SW-III and SW-IV. Later, in the paper \cite{TTW:2001} it was shown 
that they are exactly-solvable, they have the same hidden algebra $b_3 \in sl(3,R)$ and, by making some intelligent 
tricks, both those systems can be reduced to the SW-I. It implies that the SW-III and SW-IV are characterized 
by a quadratic polynomial algebra of integrals. 

\subsection{Smorodinsky-Winternitz system, Case II (1965), Holt model (1982)}

Hamiltonian \cite{Fris:1965,Wint:1966} in Cartesian coordinates has the form

\begin{equation}
\label{HHolt}
  H \ = \ -\frac{1}{2}(\,\pa_x^2 \ + \  \pa_y^2\,) \ + \ \om^2\,(x^2 \ + \ 4\,y^2) \ + \ \frac{\de}{x^2} \ ,
\end{equation}
where $\om>0$ and $\de > -\frac{1}{8}$. Usually, it is called the Smorodinsky-Winternitz system II (SW-II), 
sometimes it is also called the Holt model \cite{Holt}, see \cite{Bonatsos}. The system admits two second-order
integrals
\begin{equation}
\label{AHolt}
          {\cal I}_1 \ = \ -\pa_y^2 \ + \ 8\,\omega^2\,y^2 \ ,
\end{equation}
and
\begin{equation}
\label{BHolt}
{\cal I}_2 \ = \     2 y \pa_x^2 \ - \ 2 x\pa^2_{x\,y}  \ -  \  \pa_y   +  4 \om^2\,x^2 y -  4 \frac{\de y}{x^2}  \ ,
\end{equation}
where the commutators $[H,{\cal I}_1]=[H,{\cal I}_2]=0$.
The existence of the second order differential operator ${\cal I}_1$ trivially implies that (\ref{HHolt}) 
separates in Cartesian coordinates $(x,y)$.
Note that the integral ${\cal I}_2$ (\ref{BHolt}) is a second order differential operator, which is non-separable 
in variables $(x,y)$.
The discrete symmetry of the Hamiltonian (\ref{HHolt}) and the integral ${\cal I}_1$, written in Cartesian coordinates, 
is reflection ${Z_2}^{\oplus 2}$ ($x \rightarrow -x,\,y \rightarrow -y $). The first integral ${\cal I}_1$ is symmetric wrt the both reflections, while the second integral ${\cal I}_2$ is symmetric wrt to the first reflection $(x \rar -x)$ but it
is antisymmetric wrt the second reflection: ${\cal I}_2(x,-y)=-{\cal I}_2(x,y)$.

The commutator of integrals
\begin{equation*}
{\cal I}_{12} \ \equiv \  [{\cal I}_1,\,{\cal I}_2] \ ,
\end{equation*}
is a differential operator of third order
\begin{equation}
\label{ABHolt}
       {\cal I}_{12} \ = \ -4 \Big(\, \pa^2_x\pa_y\,-\, 8\om^2\,x y\,\pa_x \ + 
       2\big( \om^2 x^2 -\frac{\delta}{x^2}\big) \ \pa_y   \ - \ 4\,{\om}^{2}y\,\Big)\ .
\end{equation}
It apparently commutes with Hamiltonian (\ref{HHolt}), $[H, {\cal I}_{12}]=0$\,. ${\cal I}_{12}$ is 
antisymmetric wrt the second reflection: ${\cal I}_{12}(x,-y)=-{\cal I}_{12}(x,y)$. 

Syzygy is realized as
\begin{equation}
{\cal I}_{12}^2\ =\
 -16 {\cal I}_1^3 + 64  H {\cal I}_1^2 -64 H^2 {\cal I}_1 + 512 \om^2 H + 32\om^2 \big( 8\delta - 11\big) {\cal I}_1 
   + 32 \om^2 {\cal I}_2^2\ ,
\end{equation}
hence, it is given by the cubic polynomial. Double commutators are easily calculated
\begin{equation}
\label{DC1-SW2}
 [{\cal I}_{1},\,{\cal I}_{12}] \ = \ 32\,\om^2\,{\cal I}_{2} \ ,
\end{equation}
\begin{equation}
\label{DC2-SW2}
 [{\cal I}_{2},\,{\cal I}_{12}] \ = \  32 H^2  - 64 H {\cal I}_1  + 24 {\cal I}_1^2 -16\om^2 \big(  8 \delta -3\big)
 \, \ .
\end{equation}
Hence, we arrive at a quadratic algebra of integrals generated by ($ H,\,{\cal I}_1,\,{\cal I}_2,\,{\cal I}_{12}$) 
in agreement with the results obtained in \cite{Post:2011} and \cite{MPW:2013}. Note that there exists 
a certain degeneration: the rhs of (\ref{DC1-SW2}) does not depend on 
$H, {\cal I}_{1}, {\cal I}_{12}$, while the rhs of (\ref{DC2-SW2}) does not depend on 
${\cal I}_{2}, {\cal I}_{12}$.

In order to find the algebraic form of the Hamiltonian and integrals let us introduce the gauge factor
\begin{equation}
\label{Gamma-Holt}
   \Gamma \ = \ e^{-\frac{\om\,\ta_1+2\, \om\, y^2}{\sqrt{2}}}\ta_1^{\frac{1}{4}+\frac{1}{4}k\, \sqrt{1+8\, \de }}
   \quad ,  \quad k=\pm 1 \ ,
\end{equation}
and then change variables
\[
\tau_1 \ = \ x^2\quad ,\qquad y \ = \ y \ .
\]
It is the ground state eigenfunction of (\ref{HHolt}).
It leads to gauge-rotated Hamiltonian and integrals in algebraic form:
\begin{eqnarray}
\label{h2}
& h \  \equiv \ \Gamma^{-1}\,(H \,-\,E^{(k)}_{0})\,\Gamma \non
\\ 
&
 \ = \ -2\,\ta_1\,\pa^2_{\ta_1} \ - \ \frac{1}{2}\, \pa^2_{y} \ - 
  \ (\beta_k \,-\,2\,\sqrt{2}\,\om\, \ta_1  )\,\pa_{\tau_1} \ + \ 2\, \sqrt{2}\,\om\,y \,\pa_{y}  \ ,
\end{eqnarray}
here $E^{(k)}_{0}  = \frac{\om\,(2\,+\,\beta_k) }{\sqrt{2}}$ is the ground state energy and $\beta_k \equiv (2+k\,\sqrt{1+8\, \delta })$,
\begin{equation}
\label{i1-2}
  i_1 \  \equiv \ \Gamma^{-1}\, ({\cal I}_1\,-\,{\cal I}_1^{(0)})\, \Gamma
 \ = \ -\pa^2_{y} \  + \ 4\,\sqrt{2}\,\om\,y\,\pa_{y}\ ,
\end{equation}
where ${\cal I}_1^{(0)}=(2\,\sqrt{2}\,\om)$ is the lowest eigenvalue (energy) of the integral ${\cal I}_1$,  
which is the Hermite operator in $y$, and
\begin{equation}
\label{i2-2}
 i_2 \  \equiv \ \Gamma^{-1}\,{\cal I}_2
\,\Gamma =    8 \,\tau_1 \,y\,\pa^2_{\tau_1} \ - \ 4 \,\tau_1\,\pa^2_{\tau_1,y} \ + \ 4 \,\beta_k\, y\,\pa_{\tau_1}
 \ - \ (\beta_k\,-\,2\, \sqrt{2}\,\om \,\tau_1)\,\pa_{y} ,
\end{equation}
\begin{eqnarray}
\label{i12-2}
 & i_{12} \  \equiv \ \Gamma^{-1}\,{\cal I}_{12}\,\Gamma
\non \\ & \ = \   -4\Big( 4\,\tau_1\,\pa^2_{\tau_1}\,\pa_{y}\  - \ 8\,  \sqrt{2} \,\om\,\tau_1 \,y\,\pa^2_{\tau_1}\ +
   \ 2 \, \left(\,\beta_k-2 \sqrt{2}\,\om\, \tau_1\,\right)\,\pa^2_{\tau_1,y}
\\ &
 \    - \ 4\, \sqrt{2}\,\om\,\beta_k \,y\,\pa_{\tau_1} \ -
 \ \om\, \left(\sqrt{2} \,\beta_k-4\,\om\, \tau_1\right)\,\pa_{y} \Big) \ ,\non
\end{eqnarray}
see \cite{TTW:2001}. It is evident that the rhs of the double commutators $[i_1, i_{12}]$ and $[i_2, i_{12}]$ is at most the 2nd degree polynomials in $(h, i_1, i_2, i_{12})$, cf. (\ref{DC1-SW2}), (\ref{DC2-SW2}). 

In terms of the maximal affine subalgebra $b_3 \in sl(3,R)$ generators
\begin{equation*}
{\cal J}_i^{-} \ = \ \frac{\pa}{\pa \tau_i} \quad , \quad  {\cal J}_{ij}^{0} \ = 
\ \tau_i\,\frac{\pa}{\pa \tau_j} \quad ; \quad i=1,2 \ ,\ (\ta_2 \equiv y)\ ,
\end{equation*}
see (\ref{b3}), we arrive at the Lie algebraic form of the Hamiltonian and integrals:
\begin{eqnarray}
 h \ =  \ -2\,{\cal J}_{11}^{0}\,{\cal J}_{1}^{-} \ - \ \frac{1}{2}\, {\cal J}_{2}^{-}{\cal J}_{2}^{-} \ + \ 2 \sqrt{2}\,\om\,({\cal J}_{11}^{0}+{\cal J}_{22}^{0}) \ - \ \beta_k\,{\cal J}_{1}^{-} \ ,\non
\end{eqnarray}
\begin{eqnarray}
 & i_1 \ =\ -\,{\cal J}_{2}^{-}{\cal J}_{2}^{-} \ + \ 4\,\sqrt{2}\,\om\,{\cal J}_{22}^{0}  \ ,
\non
\end{eqnarray}
\begin{eqnarray}
 & i_2 = 8{\cal J}_{11}^{0}{\cal J}_{21}^{0} -4 {\cal J}_{12}^{0}{\cal J}_{1}^{-}
 + 4\beta_k {\cal J}_{21}^{0} + 2\sqrt{2}\omega {\cal J}_{12}^{0} - \beta_k {\cal J}_{2}^{-}\ , 
\non
\end{eqnarray}
\begin{eqnarray}
 & i_{12}\ = \ 4\,{\cal J}_{11}^{0}\,{\cal J}_{1}^{-}\,{\cal J}_{2}^{-}
 - 8\,\sqrt{2}\,\om\,{\cal J}_{11}^{0}\,{\cal J}_{21}^{0} \ + \ 2\,\beta_k\,
 {\cal J}_{1}^{-}\,{\cal J}_{2}^{-} \ - \ 4\,\sqrt{2}\,\om\,{\cal J}_{11}^{0}\,{\cal J}_{2}^{-}
 \\
 &
 \ - \ 4\,\sqrt{2}\,\om\,\beta_k\,{\cal J}_{21}^{0} \ - \ \sqrt{2}\,\om\,\beta_k\,
 {\cal J}_{2}^{-} \ + \ 4\,\om^2\,{\cal J}_{12}^{0}  \ .\non
\end{eqnarray}
see \cite{TTW:2001}. It implies that the SW-II system has the hidden algebra 
$sl(3,R) \in g^{(1)}$ similar to SW-I. 
Spectra of $h$ (\ref{h2}):
\begin{equation}
 \veps_{n_1,n_2} \ = \ 2\,\sqrt{2}\,\om\,( \, n_1 \,+\,n_2\,) \ ,\quad n_1,n_2=0,\,1,\,2,\ldots\ ,
\end{equation}
corresponds to the isotropic harmonic oscillator with polynomial eigenfunctions given by,
\begin{equation}
    P_{n_1,n_2}(\ta_1,\,\ta_2) \ = \ L_{n_1}^{\small(\frac{1}{2}k\,\sqrt{1+8\,\delta})}(\sqrt{2}\,\om\,\ta_1)\,
    H_{n_2}(2^{\frac{3}{4}}\,\sqrt{\om}\,y) \ ,
\end{equation}
as a product of generalized Laguerre polynomial by Hermite polynomial.

It is natural to require that the whole SW-II system - Holt model is invariant wrt reflection 
${Z_2}^{\oplus 2}$ ($x \rightarrow -x,\,y \rightarrow -y $), both the Hamiltonian and integrals. 
The Hamiltonian (\ref{HHolt}) and the first integral (\ref{AHolt}) are already ${Z_2}^{\oplus 2}$-invariant. 
In order to have the second integral ${\cal I}_2$ to be ${Z_2}^{\oplus 2}$-invariant we have to take its square
(\ref{BHolt}),
\begin{equation}
\label{I2^2}
  {\cal \hat I}_2\ = \ {\cal I}_2^2\ .
\end{equation}
It leads to a modification of the commutator of the integrals,
\[
 {\cal \hat I}_{12}\ = \ {\cal I}_{12} {\cal I}_2 + {\cal I}_2 {\cal I}_{12}\ ,
\]
and also double commutators.
{
In order to find the algebraic form of the Hamiltonian and integrals let us introduce 
the gauge factor
\begin{equation}
\label{Gamma-Holt-2}
   \Gamma_2 \ = \ e^{-\frac{\om\,\ta_1+2\, \om\, \ta_2}{\sqrt{2}}}\ta_1^{\frac{1}{4}+\frac{1}{4}k\, 
   \sqrt{1+8\, \de }}\quad ,\quad k=\pm 1 \ ,
\end{equation}
where
\[
\tau_1 \ = \ x^2\quad ,\qquad\ta_2 \ = \ y^2 \ ,
\]
cf.(\ref{Gamma-Holt}), which are invariants of the reflection group.
(\ref{Gamma-Holt-2}) is the ground state eigenfunction of (\ref{HHolt}) written in coordinates 
$\ta_1, \ta_2$. Then it has to be made the gauge rotation. As for the Hamiltonian in algebraic 
form it is, in fact, (\ref{h2}) with $y$ replaced by $\ta_2 = y^2$,
\begin{equation}
\label{h2-2}
 h_2 \ = \ -2\,\ta_1\,\pa^2_{\ta_1} \ - \ 2\,\ta_2\,\pa^2_{\ta_2}\ - 
 \ (\beta_k \,-\,2\,\sqrt{2}\,\om\, \ta_1  )\,\pa_{\ta_1}\ -\ 
 (1 - 4\sqrt{2}\,\om\,\ta_2)\,\pa_{\ta_2}\ ,
\end{equation}
as well as the first integral (\ref{i1-2}) is
\begin{equation}
\label{i1-2-2}
  i_1 \ = \ -4\ta_2\pa^2_{\ta_2} \  - \ 2(1 - 4\,\sqrt{2}\,\om\,\ta_2)\,\pa_{\ta_2}\ ,
\end{equation}
which can be recognized as the Laguerre operator. 
As for the second integral,
\begin{align}
\label{i2-2-2}
{\tilde i}_2 \ & \equiv \ \Gamma_2^{-1}\,{\cal I}^2_2 \,\Gamma_2 \vert_{\ta_1, \ta_2} \\     
&= 64\,{\ta_1}^{2} \ta_2\, {\pa^{4}_{\ta_1}}
-128\,{\ta_1}^{2} \ta_2\, {\pa^{3}_{\tau_1}\pa_{\ta_2}}
+64\,{\ta_1}^{2} \ta_2\, {\pa^{2}_{\ta_2}\pa^{2}_{\ta_1}}
\non \\ &
-32 \left( \, {\tau_1}^{2}  - 2\left( 2+\beta \right) \tau_1 \tau_2\,\right)  {\partial^{3}_{\tau_1}}
+ \left(
32\left( 2\sqrt{2}\,\omega\, \tau_2 + 1\right)\tau_1^2
- 96\left(2 + \,\beta \right) \tau_1 \tau_2\, \right) {\partial^{2}_{\tau_1}\partial_{\tau_2}}
\non \\ &
%%%%%%%%%%%%%%%%%%%%%%%%%%%%%%%%%%%
-32\left(  2\,\sqrt {2}\omega\,{\tau_1}^{2}\tau_2
- \left( 2+\,\beta \right) \tau_2\,\tau_1
\right) {\partial_{\tau_1} \partial^2_{\tau_2} }
\non \\ &
%%%%%%%%%%%%%%%%%%%%%%%%%%%%%%%%%%%
+ 8\left(
2\, \sqrt {2}\omega\,{\tau_1}^{2}
- \left( 4+3\,\beta \right)\tau_1
+ 2\,\beta\, \left( 2+\beta \right)\tau_2
\right) {\partial^{2}_{\tau_1}}
\non \\ &
%%%%%%%%%%%%%%%%%%%%%%%%%%%%%%%%%%%
-16\left(
  2\,\sqrt {2}\omega\,{\tau_1}^{2}
- 2\,\sqrt{2}\omega\, \left( 2+\beta \right)\tau_1 \tau_2
- \left( 2+\,\beta \right) \tau_1
+\,\beta\, \left( 2+\beta \right)\tau_2
\right) {\partial^{2}_{\tau_1 \tau_2}}
\non \\ &
%%%%%%%%%%%%%%%%%%%%%%%%%%%%%%%%%%%
+ 4 \left(
8\, {\omega}^{2}{\tau_1}^{2}\tau_2
- 4\,\sqrt {2} \,\omega\,\left( 2+\beta \right)\,\tau_1\,\tau_2
+  \,{\beta}^{2}\tau_2
\right)\,{\partial^{2}_{\tau_2}}
\non \\ &
%%%%%%%%%%%%%%%%%%%%%%%%%%%%%%%%%%%
+ 4\beta\left(2\,\sqrt {2}\om\,\ta_1 - \,{\beta}
\right) {\pa_{\ta_1}}
%%%%%%%%%%%%%%%%%%%%%%%%%%%%%%%%%%% 
+2\left(8\, {\om}^{2}{\tau_1}^{2}
- 4\,\sqrt{2}\,\om\, \left( 2+\beta \right)\tau_1
+ 8\,\sqrt {2}\om\,\beta\, \tau_2
+ \,{\beta}^{2}
\right){\pa_{\ta_2}}\ ,
\non
\end{align}
%%%%
which is the fourth order differential operator. New commutator ${\tilde i}_{12}=[i_1,{\tilde i}_2 ]$ is a 
fifth order differential operator,  
\begin{align}
\tilde{i}_{12} = &
  -512 \ta_1^2 \tau_2 \pa_{\tau_1}^4 \pa_{\tau_2}
  + 512 \ta_1^2 \tau_2 \pa_{\tau_1}^3 \pa_{\tau_2}^2
  + (512 \sqrt{2} \om \tau_2 - 128) \tau_1^2 \pa_{\tau_1}^4
  + 256 \tau_1 (\tau_1 - (2 \beta + 4) \tau_2) \pa_{\tau_1}^3 \pa_{\tau_2} \non \\
& + 384 \tau_1 \tau_2 (\beta + 2 - 2 \sqrt{2} \om \tau_1) \pa_{\tau_1}^2 \pa_{\tau_2}^2
  + 128 (\beta + 2) \tau_1 (4 \sqrt{2} \om \tau_2 - 1) \pa_{\tau_1}^3  \non \\
& - 64 [6 \sqrt{2} \om \ta_1^2 - 3 (\beta + 2) \ta_1 + 2 \beta (\beta + 2) \ta_2] \pa_{\ta_1}^2 \pa_{\ta_2} 
\label{i12t} 
\\
& + 64 \tau_2 [16 \om^2 \tau_1^2 - 6 \sqrt{2} \om (\beta + 2) \tau_1 + \beta (\beta + 2)] \pa_{\tau_1} \pa_{\tau_2}^2 
\non \\
& + 32 \beta (\beta + 2) (4 \sqrt{2} \om \tau_2 - 1) \pa_{\ta_1}^2
 + 32 [16 \om^2 \tau_1^2 - 6 \sqrt{2} \om (\beta + 2) \ta_1 + \beta (\beta + 2)] \pa_{\ta_1} \pa_{\tau_2} 
\non \\
& - 32 \ta_2 [8 \sqrt{2} \om^3 \ta_1^2 - 8 \om^2 (\beta + 2) \ta_1 + \sqrt{2} \om \beta (\beta + 2)] \pa_{\ta_2}^2 
\non \\
& - 16 \sqrt{2} \om [8 \om^2 \ta_1^2 - 8 \om (\beta + 2) \ta_1 + \beta (\beta + 2)] \pa_{\ta_2} \ .
\non
\end{align}
It can be checked that the hidden algebra of the system written $\ta_{1,2}$ variables is $g^{(2)}$: 
the Hamiltonian $h_2$ and the integrals $i_1, \tilde{i}_{2}, \tilde{i}_{12}$ can rewritten in terms of 
$g^{(2)}$-generators, see Section \ref{gk}.
 
In straightforward calculation the double commutators can be found in terms of 
$(h_2, i_1, {\tilde{i}}_2, \tilde{i}_{12})$:
\begin{align*}
[i_1,\tilde{i}_{12}] & = -128\, h_2^2 i_1 + 128\, h_2 i_1^2 - 32\,i_1^3
   -256\,\sqrt{2}\om\,h_2^2 - 128 \sqrt{2} \om (\beta  - 2 ) h_2 i_1
\\&
      + 64 \sqrt{2}\om (\beta-1) i_1^2 +  128 \om^2 \,\tilde{i}_2 -  
      512\om^2 (\beta -2 ) h_2 - 512 \om^2 i_1\ ,
\end{align*}
\begin{align*}
[\tilde{i}_2,\tilde{i}_{12}] & =
128 h_2^2 \,\tilde{i}_2
-256 h_2 i_1 \tilde{i}_2 
-2048 h_2^3
+ 2560 h_2^2 i_1
+ 96 i_1^2 \tilde{i}_2 -1536 h_2 i_1^2 +
384 i_1^3 +
\\ &
+ 128 h_2 \tilde{i}_{12} -96 i_1 \tilde{i}_{12}+
128\sqrt{2}\om (\beta-2)  h_2 \tilde{i}_2
-128\sqrt{2}\om (\beta-1) i_1 \tilde{i}_2 
 \\ &
 -1024\sqrt{2}\om (3\beta+1)  h_2^2
 +512\sqrt{2}\om (5\beta-2)  h_2 i_1
-768\sqrt{2}\om (\beta-1)  i_1^2 
 \\ &
 -2048\om^2 (\beta+5) (\beta-2)  h_2  +
 2048\om^2 (\beta^2-2\beta-5)  i_1 +
 4608\om^2 \tilde{i}_2 +
 64\sqrt{2}\om (\beta-1)  \tilde{i}_{12}\,.
\end{align*}
We arrive at the cubic algebra of discrete-symmetry-supported integrals.

\subsection{Quadratically superintegrable systems}

In \cite{Post:2011}, see also \cite{MPW:2013}, several (additional) quadratically superintegrable systems 
(with integrals given by the second order differential operators, at most) were considered constructively. 
All of them were characterized by the second order (quadratic) polynomial algebras of integrals. 
It was naturally conjectured that all quadratically superintegrable systems have the second order polynomial 
algebras of integrals even though the Hamiltonian is non-Hermitian or has no discrete spectra. 

\subsection{Fokas-Lagerstrom potential (1980)}

Hamiltonian, see \cite{Fokas:1980,Bonatsos},

\begin{equation}
\label{HFokas}
H \ = \ -\frac{1}{2}(\,\pa_x^2 \ + \  \pa_y^2\,) \ + \ \frac{\om^2}{2}\,\bigg(x^2 \ +
   \ \frac{1}{9}\,y^2\bigg)  \ ,
\end{equation}
where for $\om > 0$ the Hamiltonian is Hermitian and has infinite discrete spectra. 
This system is superintegrable and possesses two integrals, one of the second order
\begin{equation}
\label{AFokas}
{\cal I}_1 \ = \ -\pa_x^2 \ + \ \om^2\,x^2 \ ,
\end{equation}
and one of the third order
\begin{equation}
\label{BFokas}
{\cal I}_2 \ = \  x\,\pa_y^3 \ - \ y\,\pa_y^2\pa_x \ - \ \pa_x \pa_y \ + \ \frac{1}{3}x\,y^2\,\om^2\,\pa_y \ - \ \frac{1}{27}y^3\,\om^2\,\pa_x \ + \ \frac{1}{3}x\,y\,\om^2
 \ .
\end{equation}
The existence of the second order differential operator ${\cal I}_1$ trivially implies that (\ref{HFokas}) 
admits a separation of variables in Cartesian coordinates $(x,y)$.
The integral ${\cal I}_2$ is a third order differential operator, which is non-separable in variables $(x,y)$. 
The discrete symmetry of the Hamiltonian (\ref{HFokas}), written in Cartesian coordinates, is 
${Z_2}^{\oplus 2}$ ($x \rightarrow -x,\,y \rightarrow -y $). The integral ${\cal I}_1$ is invariant wrt reflections. 
However, the integral ${\cal I}_2$ is anti-invariant: 
\[
{\cal I}_2(x,y)=-{\cal I}_2(-x,y)=-{\cal I}_2(x,-y)={\cal I}_2(-x,-y)\ ,
\]
wrt one reflection. It must be emphasized that the second integral squared ${\cal I}_2^2$ is a differential operator 
of 6th order which is invariant under both reflections ${Z_2}^{\oplus 2}$.

The commutator of integrals
\begin{equation*}
{\cal I}_{12} \ \equiv \  [{\cal I}_1,\,{\cal I}_2] \ ,
\end{equation*}
is a differential operator of fourth order
\begin{eqnarray}
& {\cal I}_{12} \ = \  -2\,\pa_y^3\pa_x \ + \ 2 \,x\, y\, \om^2\,\pa_y^2 \ -\ 
  \frac{2}{3}\,\om^2\,y^2\,\pa_x\pa_y \ + \ 2\,x\,\om^2\,\pa_y
\\ &
\ - \ \frac{2}{3}\,\om^2\,y\,\pa_x\ \ + \ \frac{2}{27}\, x\, y^3\,\om^4 \ . \non
\end{eqnarray}

Double commutators read
\begin{equation}
\label{65}
 [{\cal I}_{1},\,{\cal I}_{12}] \ = \ 4\,\om^2\,{\cal I}_{2} \ ,
\end{equation}
\begin{equation}
\label{66}
 [{\cal I}_{2},\,{\cal I}_{12}] \ = \ 8\,{\cal I}_{1}^3 \ - \ 36\,{\cal I}_{1}^2\,H  \ + \ 48\, {\cal I}_{1}\,H^2 \ - \ 
 16\,H^3 \ + \ \frac{56}{9}\,\om^2\,{\cal I}_{1} \ - \ \frac{92}{9}\,\om^2\,H   \ .
\end{equation}
They have quite peculiar form: the first double commutator (\ref{65}) is linear in ${\cal I}_{2}$ and does not depend on 
$H, {\cal I}_{1}, {\cal I}_{12}$ while the second double commutator (\ref{66}) does not depend on ${\cal I}_{2}, {\cal I}_{12}$.
Hence, we arrive at a {\it cubic} algebra of integrals ($H,\,{\cal I}_1,\,{\cal I}_2,\,{\cal I}_{12}$).
Note there exists an algebraic relation (syzygy) involving $H,\,{\cal I}_1,\,{\cal I}_2,\,{\cal I}_{12}$ in the form 
of the fourth order polynomial,
{\small
\begin{align}
\label{67}
{\cal I}_{12}^2 & =
   32 H^3{\cal I}_1  - 48 H^2{\cal I}_1^2  + 24  H{\cal I}_1^3 - 4 {\cal I}_1^4   + 4 \om^2 {\cal I}_2^2 - 48\om^2 H^2 + 
   \frac{616}{9}\om^2 H {\cal I}_1 -\frac{200}{9}\om^2 {\cal I}_1^2 -  \frac{20}{9}\om^4\ ,
\end{align}
} cf.\cite{Bonatsos}, made from ordered monomials in ($H,\,{\cal I}_1,\,{\cal I}_2,\,{\cal I}_{12}$).

Let us introduce the gauge factor
\begin{equation}
 \Gamma \ = \ e^{-\frac{1}{2}\,\om\,x^2 - \frac{1}{6}\,\om\,y^2} \ ,
\end{equation}
which is the meaning of the ground state function of the Hamiltonian (\ref{HFokas}). 
Gauge-rotated Hamiltonian and integrals get the algebraic forms:
\begin{eqnarray}
\label{h3alg}
& h \  \equiv \ \Gamma^{-1}\,(H \,-\,E_{0})\,\Gamma
\\ & \ = \ -\frac{1}{2}\,\pa^2_{x} \ - \ \frac{1}{2}\, \pa^2_{y} \ + \ \om\,x\,\pa_{x} \ +\ 
 \frac{1}{3}\,\om\,y\,\pa_{y}   \ ,\non
\end{eqnarray}
which is the sum of two Hermite operators and $E_{0}=\frac{2}{3}\, \om\,$ 
is the ground state energy of the Hamiltonian (\ref{HFokas}), and
\begin{equation}
\label{i13alg}
 i_1 \ \equiv \ \Gamma^{-1}\, (\,{\cal I}_1\,-\,{\cal I}_1^{(0)})\, \Gamma
 \ = \ -\pa^2_{x} \  + \ 2\,\om\,x\,\pa_{x}\ ,
\end{equation} 
where ${\cal I}_1^{(0)}=\om$ is the lowest eigenvalue of the integral ${\cal I}_1$, and
\begin{eqnarray}
\label{i23alg}
& i_2 \  \equiv \ \Gamma^{-1}\,{\cal I}_2
\,\Gamma
\\ & \ =\ -y\,\pa^2_{y}\,\pa_{x} \ + \ x\,\pa^3_{y} \ + \ 
   \big(\,\frac{2 \, \om}{3}\,y^2-1\big)\pa^2_{xy}\ + \ \frac{2}{27}\, \om\,y\,  
   \left(9-2\, \om\,y^2 \right)\,\pa_{x}\ .\non
\end{eqnarray} 
Finally,
\begin{eqnarray}
&  i_{12}\  \equiv \ \Gamma^{-1}\,{\cal I}_{12}\,\Gamma
\\ 
& \ = \ -2\,\pa_{x} \pa^3_{y} \ + \ 2\,\om\,x\,\pa^3_{y} \ + 
   \ 2\,\om\,y\,\pa_{x}\,\pa^2_{y} \ + \ \frac{2}{3}\, \om\,\left(3-2 \, \om\,y^2 \right)\,\pa^2_{xy} \ +\ 
   \frac{4}{27}\,  \om^2\,y\, \left(2 \, \om\,y -9\right)\,\pa_{x}
\ ,\non
\end{eqnarray}

In terms of the $g^{(3)}$ generators, see Section \ref{gk}, presented also in \cite{vieyra2023wolfes} (eqs.(20)-(22)), 
we arrive at the (Lie)-algebraic form of the Hamiltonian (69) and integrals (70)-(71):
\begin{eqnarray}
 h \ =  \ -\frac{1}{2}( {\cal R}_0\,{\cal R}_0 \ + \ {\cal J}^1\,{\cal J}^1  ) \ + \ 
 \frac{\om}{3}\,\big({\cal J}^2_0 \,+\,{\cal J}^3_0 \big) \ ,\non
\end{eqnarray}
\begin{eqnarray}
 & i_1 \ =  \  - \,{\cal R}_0\,{\cal R}_0 \ + \ \frac{2\,\om}{3}\,{\cal J}^3_0 
  \ ,\non
\end{eqnarray}
\begin{eqnarray}
 & i_2 \  = \  {\cal R}_1\,{\cal J}^1\,{\cal J}^1 \ + \ {\cal T}_0^{(3)} \ + \ \frac{2\,\om}{3}\,{\cal R}_2\,{\cal J}^1 \ + 
 \ \frac{2\,\om}{3}\,{\cal R}_1 \ - \ \frac{4\,\om^2}{27}\,{\cal R}_3
 \ ,\non
\end{eqnarray}
where we made the identifications in Section III, see also\cite{vieyra2023wolfes}: $x=u$ and $y=r$. Hence, $g^{(3)}$ is 
the hidden algebra of the Fokas-Lagerstrom model.

Spectra of the algebraic Hamiltonian $h$ (\ref{h3alg}):
\begin{equation}
\veps_{n_1,n_2} \ = \ \frac{1}{3}\,\om\,(3\,n_1 \,+\,n_2) \ ,
\end{equation}
where $n_1,n_2=0,\,1,\,2,\ldots$ are quantum numbers, which corresponds to the 2D anisotropic harmonic oscillator with frequency ratio $3:1$, its eigenfunctions are factorized, 
\begin{equation}
P_{n_1,n_2}(x,\,y) \ =\ H_{n_1}(\,\sqrt{\om}\,x)\,H_{n_2}(\,\sqrt{\om/3}\,y) \ ,
\end{equation}
to the product of two Hermite polynomials.

By imposing a condition of ${Z_2}^{\oplus 2}$ invariance on the Hamiltonian and both integrals 
of the Fokas-Lagerstrom model we arrive at their algebraic forms, reveal hidden algebra and eventually find a polynomial
algebra of integrals of ${Z_2}^{\oplus 2}$ invariant nature. 
The simplest way to realize this condition is to introduce ${Z_2}^{\oplus 2}$ invariants 
as new variables,
\[
\ta_1=x^2\ ,\ \ta_2=y^2 \ ,
\]
instead of the Cartesian coordinates and then require the algebraic forms of the Hamiltonian 
and both integrals exist, if they do.  Remarkably, the Hamiltonian (\ref{h3alg}) and 
the first integral (\ref{i13alg}) remain algebraic in $\ta_1,\ta_2$ coordinates: 
after the change of variables,
\begin{equation}
\label{h3alg-inv}
 h_{invariant} \ =\ -2\ta_1\pa^2_{\ta_1} \  - \ \Big(1 - 2\,\om\,\ta_1\Big)\,\pa_{\ta_1}
 -2\ta_2\pa^2_{\ta_2} \  - \ \Big(1 - \frac{2}{3}\,\om\,\ta_2\Big)\,\pa_{\ta_2}\ ,
\end{equation}
which is the sum of two Laguerre operators, and
\begin{equation}
\label{i13alg-inv}
 i_1 \ = \ -4\ta_1\pa^2_{\ta_1} \  - \ 2(1 - 2\,\om\,\ta_1)\,\pa_{\ta_1}\ .
\end{equation} 
However, in order to get the second integral in the algebraic form we have to square the integral 
$i_2$ (\ref{i23alg}) and then change variables $(x,y)$ to $\ta_{1,2}$:
{\footnotesize  %%%%%%%%%%%%%%%%%%%%%%%%%%%%%
\begin{equation}
%%%%%%% Reorganized and checked on Oct 08, 2025
  \tilde{i}_2 \equiv i_2^2\ =\ 
% Total order 6
 64 \ta_1 \ta_2^3 \pa_{\ta_1}^2 \pa_{\ta_2}^4 \ -\ 128 \ta_1 \ta_2^3 \pa_{\ta_1} \pa_{\ta_2}^5 \ 
 +\ 64 \ta_1 \ta_2^3 \pa_{\ta_2}^6 
\end{equation}
%\\ &
\begin{align*}
% Total order 5
& - \frac{64}{3} \ta_1 \tau_2^2 (2 \om \ta_2 - 15) \pa_{\tau_1}^2 \pa_{\tau_2}^3 \ 
 +\ \frac{32}{3} \ta_2^2 (4 \om \ta_1 \ta_2 - 75 \ta_1 + 3 \ta_2) \pa_{\ta_1} \pa_{\ta_2}^4 \
 +\ 32 \ta_2^2 (15 \ta_1 - \ta_2) \pa_{\ta_2}^5 \ 
\\ &
% Total order 4
 +\ \frac{16}{27} \ta_1 \ta_2 (20 \om^2 \ta_2^2\ -\ 270 \om \ta_2 + 459) \pa_{\ta_1}^2 \pa_{\ta_2}^2 \
 -\ \frac{32}{27} \left(
 4\om^2 \ta_2^3\ta_1\,-\,18 \om (10\ta_1\ta_2^2 + \ta_2^3)\,+\,810\ta_2\,\ta_1 - 135\ta_2^2
 \right) \pa_{\ta_1} \pa_{\ta_2}^3 
\\ &
 +\ \frac{16}{3} \ta_2 (2 \om \ta_2^2 + 135 \ta_1 - 33 \ta_2) \pa_{\ta_2}^4 \ 
% Total order 3
 -\ \frac{16}{81} \ta_1 (8 \om^3 \tau_2^3 - 150 \om^2 \tau_2^2 + 459 \om \tau_2 - 81) \pa_{\tau_1}^2 \pa_{\ta_2} \ 
\\ &
 -\ \frac{8}{27} \left(20 \om^2(3\ta_1 - \ta_2)\,\ta_2^2\ 
 -\ 54 \om (13\ta_1 - 5\ta_2)\,\ta_2\ +\ 405\ta_1 - 459\ta_2 \right) \pa_{\ta_1} \pa_{\ta_2}^2 
\\ &
 -\ \frac{8}{27} \left(4 \om^2 \ta_2^3 - 126 \om \ta_2^2 - 405 \ta_1 + 567 \ta_2 \right) \pa_{\ta_2}^3 \,
% Total order 2
+ \frac{8}{729}  \om (8 \om^3 \ta_2^3 - 180 \om^2 \ta_2^2 + 810 \om \ta_2 - 243)\ta_1 \pa_{\ta_1}^2 \,
\\ &
 -\ \frac{8}{81} \left(
 8\om^3\ta_2^3 + 6 (27\ta_1\tau_2 - 25\ta_2^2)\om^2
 - 27 (9\ta_1- 17\ta_2)\om  - 81 \right) \pa_{\ta_1} \pa_{\tau_2} \ 
\\ &
 -\ \frac{4}{9} \left(4 \om^2 \ta_2^2 - 36 \om \ta_2 + 27 \right) \pa_{\ta_2}^2 \ 
%%\\ &
% Total order 1
  +\ \frac{4}{729} \om \left(8 \om^3 \ta_2^3 - 180 \om^2 \ta_2^2 + 
  (- 324 \ta_1 + 810 \ta_2)\om -243\right) \pa_{\ta_1} \ ,
\end{align*}
}
which is the sixth order differential operator, and then calculate the commutator
{\footnotesize 
\begin{equation}
  %%%%% reorganized and checked
\tilde{i}_{12}\ \equiv \ [i_1, \tilde{i}_2]\  
 =\ 512 \ta_1 \ta_2^3 \pa_{\ta_1}^2 \pa_{\ta_2}^5 \,
 - 512 \ta_1 \ta_2^3 \pa_{\ta_1} \pa_{\ta_2}^6 \,
 + 128 \ta_2^3 (2 \om \ta_1 - 1) \pa_{\ta_2}^6 \,
\end{equation} 
\begin{align*}
&
 - \frac{640}{3} \tau_1 \tau_2^2 (2 \om \tau_2 - 5) \pa_{\ta_1}^2 \pa_{\ta_2}^4 \,
 - 256 \ta_2^2 (15 \ta_1 - \ta_2) \pa_{\ta_1} \pa_{\ta_2}^5 \,
 + 960 \ta_2^2 (2 \om \ta_1 - 1) \pa_{\ta_2}^5 \,
\\ &
+ \frac{1280}{27} \tau_1 \tau_2 (4 \omega^2 \tau_2^2 - 45 \omega \tau_2 + 81) \partial_{\tau_1}^2 \partial_{\tau_2}^3 \,
- \frac{320}{3} \tau_2 (2 \omega \tau_2^2 + 54 \tau_1 - 15 \tau_2) \partial_{\tau_1} \partial_{\tau_2}^4 \,
+ 1440 \tau_2 (2 \omega \tau_1 - 1) \partial_{\tau_2}^4 \,
\\ &
- \frac{160}{27} \ta_1 (8 \om^3 \ta_2^3 - 120 \om^2 \ta_2^2 + 324 \om \ta_2 - 81) \pa_{\ta_1}^2 \pa_{\ta_2}^2 \,
+ \frac{320}{27} \left(8 \om^2 \ta_2^3 - 90 \om \ta_2^2 - 81 \ta_1 + 162 \ta_2 \right) \pa_{\tau_1} \pa_{\ta_2}^3 \,
\\ &
+ \frac{32}{81} \om \ta_1 (16 \om^3 \ta_2^3 - 300 \om^2 \ta_2^2 + 1080 \om \ta_2 - 405) \pa_{\ta_1}^2 \pa_{\ta_2} \,
 -\ \frac{80}{27} \left(8 \om^3 \ta_2^3 - 120 \om^2 \ta_2^2 + 324 \om \ta_2 - 81 \right) \pa_{\ta_1} \pa_{\ta_2}^2
\\ &
+ 240 (2 \om \ta_1 - 1) \pa_{\ta_2}^3 \,
- \frac{32}{729} \om^2 \ta_1 (8 \om^3 \ta_2^3 - 180 \om^2 \ta_2^2 + 810 \om \ta_2 - 405) \pa_{\ta_1}^2 \,
\\ &
 +\ \frac{16}{81} \om (16 \om^3 \ta_2^3 - 300 \om^2 \ta_2^2 + 1080 \om\ta_2 - 405) \pa_{\ta_1} \pa_{\ta_2}\
 -\ \frac{16}{729} \om^2 (8 \om^3 \ta_2^3 - 180 \om^2 \ta_2^2 + 810 \om \ta_2 - 405) \pa_{\ta_1} \ ,
\end{align*}
}
which is the seventh order differential operator. Their hidden algebra is $g^{(3)}$: 
$(h,i_1,\tilde{i}_2,\tilde{i}_{12})$ can be rewritten in terms of $g^{(3)}$ generators, 
they preserve the infinite flag ${\cal P}^{(3)}$, see Section \ref{gk}.

Syzygy
\begin{align*}
\tilde{i}_{12}^2 &=
 128\, h^3\, i_1\, \tilde{i}_2
 -192\,h^2\,i_1^2\, \tilde{i}_2
 + 96\,h\,i_1^3\, \tilde{i}_2
 -16\, i_1^4\, \tilde{i}_2
 -64 \, h^3\,  \tilde{i}_{12}
+ 192 \, h^2\,i_1\, \tilde{i}_{12}
-144 \, h\, i_1^2\, \tilde{i}_{12}
\\ &
+ 32\, i_1^3 \, \tilde{i}_{12}
-768\,h^6
+ 1536\,h^5\,i_1
-1152\, h^4\,i_1^2
+ 384\, h^3\, i_1^3
-48\, h^2\, i_1^4
+ 16 \,\om^2 \,\tilde{i}_2^2
\\ &
+ 128\, \om\, h^3\,\tilde{i}_2
-128\,\om\, h^2i_1\,\tilde{i}_2
+ 32\,\om\, h\,i_1^2\,\tilde{i}_2
+ 64\, \om\, h^2\,\tilde{i}_{12}
-32\,\om\, h\,i_1\, \tilde{i}_{12}
\\&
-1536\,\om\, h^5
+2816\, \om\, h^4\,i_1
-1920\,\om\,h^3\, i_1^2
+576\,\om\, h^2\, i_1^3
-64\,\om\,h \, i_1^4
-1664\,\om\,^2\,h^2\,\tilde{i}_2
\\ &
+(22720/9)\,\om\,^2\,h\,i_1\,\tilde{i}_2
-(7616/9)\,\om\,^2\,i_1^2\,\tilde{i}_2
-(7904/9)\,\om\,^2\,h\, \tilde{i}_{12}
+ (5312/9)\, \om^2\,i_1 \tilde{i}_{12}
\\ &
-(6400/3)\,\om\,^2\,h^4
+(1792/9)\,\om\,^2\,h^3\,i_1
+ (7616/3)\,\om^2\,h^2\,i_1^2
-(4864/3)\,\om\,^2\,h\,i_1^3 
\\ &
+ (2560/9)\,\om^2\,i_1^4
+(2560/9)\,\om^3\,h\,\tilde{i}_2
-(41984/9)\,\om\,^3\,h^3
+(13312/3)\,\om\,^3\,h^2\,i_1
\\ &
-(9472/9)\,\om\,^3\,h\,i_1^2
-(24064/9)\,\om\,^4\,\,\tilde{i}_2
+(201728/27)\, \om^4\,h^2
-(1030144/81)\,\om^4\,h\,i_1
\\ &
+ (358400/81)\,\om^4\,i_1^2
-(151552/81)\, \om^5\,h \ ,
\end{align*}
cf.(\ref{67}), 
is the fifth order polynomial made from ordered monomials in $(h,i_1,\tilde{i}_2,\tilde{i}_{12})$. 
 
Double commutators
 \begin{align*}
[i_1, \tilde{i}_{12}] &\ =\ 
  64 h^3\, i_1
- 96 h^2\, i_1^2
+ 48 h\, i_1^3
- 8 i_1^4
+ 64 \om h^3
- 64 \om h^2\, i_1
+ 16 \om h\, i_1^2 \\
&- 64 \om^2 h^2
+ \frac{992}{9} \om^2 h\, i_1
- \frac{352}{9} \om^2 i_1^2 + \frac{128}{9} \om^3 h
+ 16 \om^2 \tilde{i}_2\ ,
\end{align*}
and
{\small
\begin{align*}
[\tilde{i}_2, \tilde{i}_{12}] &= 
  -768 h^5 
  + 1152 h^4 i_1 
  - 576 h^3 i_1^2 
  + 96 h^2 i_1^3 
  \\& 
  -1408 \om h^4 
  + 1920 \om h^3 i_1 
  - 64 h^3 \tilde{i}_2 
  - 864 \om h^2\, i_1^2 
  + 192 h^2\, i_1\, \tilde{i}_2 
  + 128 \om h\, i_1^3 
  - 144 h\, i_1^2\, \tilde{i}_2 
  + 32 i_1^3\, \tilde{i}_2 
  \\ &
  + \frac{12928}{9} \om^2 h^3 
  - \frac{9920}{3} \om^2 h^2\, i_1 
  + 64 \om h^2\, \tilde{i}_2 
  - 96 h^2\, \tilde{i}_{12} 
  + 2432 \om^2 h\, i_1^2 
  + 144 h\, i_1\, \tilde{i}_{12} 
  - 32 \om h\, i_1\, \tilde{i}_2  
  \\ &
  - \frac{5120}{9} \om^2 i_1^3 
  - 48 i_1^2\, \tilde{i}_{12} 
%  \\ & 
  + \frac{256}{3} \om^3 h^2 
  + \frac{256}{9} \om^3 h\,i_1 
  - \frac{7904}{9} \om^2 h\,\tilde{i}_2 
  + 16 \om h\,\tilde{i}_{12} 
  + \frac{5312}{9} \om^2 i_1\,\tilde{i}_2 
\\ & 
  + -\frac{10240}{81} \om^4 h 
  + \frac{10240}{81} \om^4 i_1 
  - \frac{1504}{9} \om^2 \tilde{i}_{12} \ ,
\end{align*}
}
define {\it quintic} algebra of integrals - invariant (implicitly) wrt discrete group of transformations 
${Z_2}^{\oplus 2}$, cf.(\ref{65})-(\ref{66}).

\section{$3$-body Calogero or $A_2$ rational problem, singular case ($\om=0$)}
\label{Calogero3body}

The 3-body/$A_2$-rational Calogero model~\cite{Calogero:1969} is a
maximally superintegrable system, {\it i.e.} there exist five independent 
integrals of motion, including the Hamiltonian. 
This model describes 3 interacting bodies of equal masses, $m=1$, on 
the line with pairwise interactions. The Hamiltonian of this 
one-dimensional system is given by
\begin{equation}
\label{HCal}
 {\cal H}_{\rm Cal}\ =\
 -\frac{1}{2} \sum_{k=1}^{3}\bigg[\,{\frac {\pa^{2}}{\pa {{x}_k}^{2}}} + \om^2 x_k^2\bigg]
 \ +\ g\sum_{k<l}^{3}\frac{1}{(x_{k} - x_{l})^2}\ ,
\end{equation}
where $\om$ is the frequency of a confining harmonic oscillator
interaction, and $g=\nu(\nu-1)>-\frac{1}{4}$ is a
coupling constant. Exact solvability of this system was demonstrated explicitly
in~\cite{RT:1995}.   

The particular case of the so-called {\it pure} or singular $A_2$-rational system
($\om=0$) is characterized by a cubic polynomial algebra of
integrals~\cite{vieyra2023wolfes}: if the Center-of-Mass (CM) motion 
is separated out, the relative motion is two-dimensional, and maximal
superintegrability implies the existence of 3 algebraically
independent integrals. This separation is achieved if we make the
change of variables $(x_1,x_2,x_3) \to (y_1,y_2,Y)$, where
$y_i=x_i-\frac{1}{3}Y, i=1,2$ are relative coordinates and $Y$ is the CM 
coordinate, $Y=\sum_{k=1}^3x_k$.

In general, an algebraic form of the Hamiltonian of relative motion can be found 
if we make change of relative variables to symmetric, translation-invariant coordinates 
and perform a gauge transformation with the ground state eigenfunction. 
This change of variables to symmetric polynomials of degrees 2 and 3 
(the degrees of the invariant polynomials of the Weyl group, $W(A_2)$) 
$(y_1,y_2) \to (x,y)$ is the following:
\begin{equation}
\label{xy-variables}
        x =-(y_1^2+ y_2^2 + y_1 y_2)\quad ,\quad y = - y_1 y_2 (y_1 + y_2)\ .
\end{equation}
Finally, with a gauge transformation at $\om=0$,
\[
   h\ \equiv \ -2\ \De^{-\nu}\, {\cal H} \, \De^{\nu}\, ,
\]
where $\De$ is Vandermonde determinant, $\De=\prod_{i<j}^3|x_i-x_j|$, 
in Cartesian coordinates we obtain the Hamiltonian in algebraic form 
(the factor -2 is introduced for convenience). The change of variables
(\ref{xy-variables}) brings the Hamiltonian to the form of a
differential operator with polynomial coefficients:
\begin{equation}
\label{hA2rat}
   h \equiv h_{A_2}^{(r)}(x,y)\ =\
   x \frac{\pa^2}{\pa x^2}\ +\ 3 y \frac{\pa^2}{\pa x \pa y}\ -\ 
   \frac{1}{3} x^2 \frac{\pa^2}{\pa y^2}\ +\ (1+3\nu) \frac{\pa}{\pa x} \ .
\end{equation}
It is known that there exists a 2nd order integral, associated with the separation
of variables in relative polar coordinates (see \cite{Calogero:1969} and 
also \cite{TTW:2009}). This gauge rotated integral in $(x,y)$ coordinates  
takes the algebraic form:
\begin{equation}
\label{xA2rat}
    i_1 \equiv x^{(r)}_{A_2}(x,y)\ =\
     \frac{1}{3}\left( 4\,{x}^{3} + 27\,{y}^{2} \right) {\frac {\pa^{2}}{\pa {y}^{2}}}
     + 9\,y \left(1 + 2\,\nu \right) {\frac {\pa }{\pa y}}\, .
\end{equation}
An extra, 3rd order integral for this model in the form presented below was found 
in \cite{ST:2015} as a degeneration of one for the general elliptic 3-body Calogero model. 
It is given by the following expression:
\begin{align}
\label{kA2rat}
 i_2 \equiv {k}_{A_2}^{(r)}(x,y)\ &=\
 y{\frac {\pa^{3}}{\pa {x}^{3}}}
 - \frac{2}{3} \,{x}^{2}{\frac {\pa ^{3}}{\pa {x}^{2} \pa y}}
 - x y {\frac {\pa^{3}}{\pa x \pa {y}^{2}}}
 - \left( {y}^{2} + {\frac {2}{27}\,{x}^{3}} \right) {\frac {\pa^{3}}{\pa {y}^{3}}} \\ &
 - \frac{2}{3}\,x \left( 2+3\,\nu \right) {\frac {\pa^{2}}{\pa x\pa y}}
 - y \left( 2+3\,\nu \right) {\frac {\pa^{2}}{\pa {y}^{2}}}
 - \frac{2}{9} \left( 2 + 3\,\nu \right)  \left( 1+3\,\nu \right) {\frac {\pa}{\pa y}} 
\ . \non
\end{align}
It can be shown that all three operators $h, i_1, i_2$ can be rewritten in terms 
of the $g^{(2)}$-algebra generators, see Section \ref{gk}, where the positive root/raising 
generator $J^4_N$ is not involved. 
Hence, the $g^{(2)}$ algebra is the hidden algebra of the model.

By construction
\[
 [h,i_1] = [h,i_2] = 0\ ,
\]
while the commutator
\[
 [i_1,i_2] \equiv i_{12}\ ,
\]
is an 4th order differential operator with polynomial coefficients, which
commutes with the Hamiltonian, $[h,i_{12}]=0$.  Note that the operator $i_{12}$ is algebraically 
independent: $i_{12}$ can {\it not} be rewritten in terms of the operators $h$ and $i_1, i_2$ unlike 
$i^2_{12}$.
It is worth noting that the 2nd order integral $i_1$ is defined ambiguously,
\[
   i_1 \rar i_1 + A\,h\ ,
\]
where $A$ is an arbitrary parameter; this transformation preserves the second degree of the differential 
operator $i_1$, see (\ref{xA2rat}).

The double commutators $[i_1,i_{12}], [i_2, i_{12}]$ are 5th  and 6th order differential operators with
$(x,y)$-dependent polynomial coefficient functions, respectively.
Their rhs can be rewritten as very simple quadratic/cubic
polynomials in $h, i_1, i_2, i_{12}$\,, respectively:
\[
      [i_1, i_{12}]\ =\
        36\,i_1 i_2\ -\ 18 i_{12}\ -\ 81 (1-4 \nu^2) i_2\ ,
\]
\[
      [i_{2}, i_{12}]\ =\ -(8/3) h^3\ -\ 18 i_{2}^2\ ,
\]
where the last commutator is defined unambiguously. The conclusion is
that the operators $h, i_1, i_2, i_{12}$ satisfy
the following relations:
\[
 [h,i_{1}]\ =\ 0\ ,\ [h,i_{2}]\ =\ 0\ ,
 \ [i_{1},i_{2}]\ =\ i_{12}\ ,\ [h,i_{12}]\ =\ 0\ ,
\]
\[
 [i_1,i_{12}]\ =\ 36\,i_1 i_2\ -\ 18\,i_{12}\ -\ 81\,(1-4 \nu^2)\,i_2\ ,
\]
\begin{equation}
\label{cubic-PA}
 [i_2, i_{12}]\ =\ -(8/3) h^3\ -\ 18 i_{2}^2\ .
\end{equation}
Hence, we arrive at {\it cubic} algebra of integrals of 3-body rational Calogero model at $\om=0$.

It is worth emphasizing that the $\nu$ dependence appears in the double commutator $[i_1,i_{12}]$ {\it only}.  
Hence, we arrive at {\it cubic} one-parametric polynomial algebra of integrals (\ref{cubic-PA})
generated by the elements $(h, i_1, i_2, i_{12})$. In representation by differential operators 
this polynomial algebra is subalgebra of the algebra $g^{(2)}$. Note that syzygy can be calculated: 
it has the form 
\begin{equation}
\label{A3-syzygy}
 i_{12}^2\ =\ \frac{16}{3}\,h^3\,i_1\ +\ 24\,(2\nu -3)\,h^3\ +\ 36\,i_1\,i_2^2\ 
 +\ 81\,(4\nu^2-9)\,i_2^2\ -\ 36\,i_2 \,i_{12} \ ,
\end{equation}
being a polynomial of degree 4 in $(h, i_1, i_2, i_{12})$.

This cubic algebra of integrals (\ref{cubic-PA}) is alternative to the {\it quartic} three-parametric 
polynomial algebra of integrals of the $G_2$ rational model of which the pure $A_2$ rational system 
is a particular case at $\la=0$ and $\om=0$ (see \cite{vieyra2023wolfes}). It must be emphasized that 
for $\om \neq 0$ in (\ref{HCal}) the first integral (\ref{xA2rat}) remains while the second integral 
exists but it becomes a differential operator of degree six, similar to one for $G_2$ rational 
superintegrable model, see \cite{vieyra2023wolfes}, at $\la=0$.
The algebra of integrals is modified becoming the quartic polynomial algebra \cite{vieyra2023wolfes}, 
see next Section.

\section{$3$-body Wolfes model or $G_2/I_6$ rational problem, singular case ($\om=0$)}
\label{Wolfes3body}

The 3-body Wolfes/$G_2/I_6$-rational model~\cite{Wolfes:1974} is a
maximally superintegrable system, {\it i.e.} there exist five independent
integrals of motion, including the Hamiltonian.
This model describes 3 interacting bodies, marked by coordinates $(x_1,x_2,x_3)$,
of equal masses, $m=1$, on the line with pairwise and three-body interactions.
The Hamiltonian is given by
\[
{\cal H}_{\rm Wolfes}\ \equiv\ H_{G_2/I_6}^{(\rm rational)}\ =
\]
\begin{equation}
\label{HWol}
-\frac{1}{2} \sum_{k=1}^{3}\bigg[\,{\frac {\pa^{2}}{\pa {{x}_k}^{2}}} + \om^2 x_k^2\bigg]
+ g_s\sum_{k<l}^{3}\frac{1}{(x_{k} - x_{l})^2}
+ 3 g_l\sum_{ k<l ,\ k,l \neq m}^{3} \frac{1}{(x_{k} + x_{l}-2 x_{m})^2}\ ,
\end{equation}
where $\om$ is a frequency and $g_s=\tilde\nu(\tilde\nu-1) > -
\frac{1}{4}$, $g_l=\tilde\mu (\tilde\mu -1) > - \frac{1}{4}$ are coupling constants
associated with the two-body and three-body interactions, which are defined
by {\it short} and {\it long} roots of the $G_2$-root space, where Weyl group $W({G_2})$ acts.
If $g_l=0$ (or equivalently, \hbox{$\tilde\mu=0,1$}) the Wolfes model is reduced to the celebrated
3-body Calogero model (or, equivalently, $A_2$-rational integrable system) of the Hamiltonian
reduction, see for discussion \cite{OP:1983}), and Section \ref{Calogero3body}.
Exact solvability of this system was demonstrated explicitly in~\cite{RTC:1998},
\begin{equation}
\label{G2-spectrum}
  E_{n_1,n_2}\ =\ 2\om (n_1 + 3 n_2)\ + \
  \frac{3}{2}\,\om\, (1 + 2 \tilde \nu + 2\tilde \mu)\ ,
\end{equation}
where $n_{1,2}=0,1,2,\ldots $ are quantum numbers. This spectrum corresponds
to the harmonic oscillator with frequency ratio 1:3 with the ground state energy
\[
  E_0 = E_{0,0} = \frac{3}{2}\,\om\, (1 + 2 \tilde \nu + 2\tilde \mu) \ .
\]
If $\om=0$ the $G_2$ rational model defined by (\ref{HWol}) is called {\it singular} (or pure) 
$G_2$ rational model, in this Section namely this model will be considered.  

If CM motion in (\ref{HWol}) is separated out, the relative motion becomes two-dimensional,
naturally parametrized by polar coordinates $(r, \varphi)$ \cite{Wolfes:1974}:
the model becomes a particular case of the TTW system \cite{TTW:2009} with index $k=3$,
its Hamiltonian is
\begin{equation}
\label{Hk=3}
 H_3 (r,\varphi;\om, \al, \beta)\ =\ -\pa_r^2 -
 \frac{1}{r}\pa_r - \frac{1}{r^2}\pa_{\varphi}^2  + \om^2 r^2 +
 \frac{9\al}{r^2 \cos^2 {3 \varphi}} + \frac{9\beta}{r^2 \sin^2 {3 \varphi}}\ ,
\end{equation}
where $\al, \beta > - \frac{1}{36}$ are parameters related to $g_{s,l}$ in (\ref{HWol}).
Evidently, the Hamiltonian $H_{3}$ (\ref{Hk=3}) is invariant with respect to the dihedral $I_6$ group
transformations. Its ground state eigenfunction is given by
\[
  \Psi_0\ =\ \De^{\tilde\nu}\De_1^{\tilde\mu}\,e^{-\frac{\om}{2} r^2} \ ,
\]
where $\De=\prod_{i<j}^3|x_i-x_j|$ is Vandermonde determinant in $x$-variables
and $\De_1=\prod_{i<j; \ i,j\neq k}|x_i+x_j-2x_k|$.

Now let us introduce new coordinates
\begin{equation}
\label{uv-variables}
        u =-(y_1^2+ y_2^2 + y_1 y_2)=r^2,\qquad v = [y_1 y_2 (y_1 + y_2)]^2 \propto r^6 \cos 6 \phi\ ,
\end{equation}
see \cite{RTC:1998}, where $y_i=x_i-\frac{1}{3}Y,\ i=1,2$ 
\footnote{Sometimes they are called the Perelomov coordinates.} 
and $Y=x_1+x_2+x_3$. It is easy to check that $u,v$ are
the basic invariants of $I_6$ dihedral group (of minimal degrees). Making a gauge rotation,
\[
   h_3\ \equiv \ -2\ {\Psi_0}^{-1}\ ({H_3}-E_0)\ \Psi_0\ ,
\]
and changing variables to $(u,v)$, we arrive at extremely simple algebraic Hamiltonian:
\[
  h_3\ =\ u \frac{\pa^2}{\pa u^2}\ +\
    6 v \frac{\pa^2}{\pa u \pa v}\ -\ \frac{4 }{3} u^2 v \frac{\pa^2}{\pa v^2}\ +\
\]
\begin{equation}
\label{hG2rat}
 (1+3\nu + 6 \la) \frac{\pa}{\pa u}\ -\ \left(\frac{2}{3} + 4\la \right) u^2\frac{\pa}{\pa v}
 -\ 4\,\om\,u\,\frac{\pa}{\pa u} - 12\,\om\,v\,\frac{\pa}{\pa v}\ ,
\end{equation}
where the parameters $\la$ and $\nu$ are related to the coupling constants $g_{s,l}$ in (\ref{HWol}) as
\[
  \la = \frac{1}{3} \tilde\mu\ ,\ \nu=\tilde\nu + \frac{1}{3}\tilde\mu\ ,
\]
see \cite{RTC:1998,Turbiner:2005,vieyra2023wolfes}. For any values of the parameters $\la, \nu, \om$
the operator $h_3$ can be rewritten in generators of the algebra $g^{(s)}$ with any $s \geq 2$. At $\om=0$
the last two terms in (\ref{hG2rat}) vanish. 

It is well known \cite{OP:1983} that the $G_2/I_6$-rational model (\ref{Hk=3}) is super-integrable, 
it has two integrals of motion, see for discussion \cite{TTW:2009}: one integral is of the second 
order (due to separation of variables in polar coordinates) and another one is of the sixth order 
(inspired by the Hamiltonian reduction method \cite{OP:1983}). 
It was shown in \cite{TTW:2009} that by making the gauge rotation of the integrals with the ground 
state function  $\Psi_0$ as a gauge factor we arrive at integrals of the TTW model at $k=3$ in  algebraic form: 
one of second order, and another one of sixth order 
\footnote{Note that in \cite{TTW:2009} the algebraic forms of the Hamiltonian and integrals 
are presented in variables $t = r^2, u = r^6 \sin^2 3\phi$.}, hence, to the form of differential operators 
with polynomial coefficients in $(t,u)$ variables. If rewritten in coordinates $(u, v)$ (\ref{uv-variables}),  
the second order integral takes the following form
\begin{equation}
\label{i1-uv}
 i_1\ =\ \frac{4}{3}\,v \left( 4\,{u}^{3}+27\,v \right)
 {\frac {\pa ^{2}}{\pa {v}^{2}}}\ +\ \frac{4}{3} \left(2 (6\la +1) \,{u}^{3}
 +   27\,( 2\la + \nu\,+1)\,v \right) {\frac {\pa}{\pa v}} \ ,
\end{equation}
which does not depend on $\om$, see  \cite{vieyra2023wolfes}. Let us not that for the particular case $\om=0$, 
discussed in  \cite{vieyra2023wolfes}, the sixth-order integral has the form 
\footnote{For the explicit expression of $i_2$ see \cite{vieyra2023wolfes}, Appendix A.}.
\begin{equation}
\label{i2-uv}
      i_2\ =\ ({{k}_{A_2}^{(r)}})^2 (u,v)\  +\ \sum_{p=1}^{5} \la^{p} {k}^{(p)}(u,v) \ ,
\end{equation}
where  the first term is the square of the cubic integral of the $A_2$-rational integrable model 
$k_{A_2}^{(r)}(x,y)$ (\ref{kA2rat}), and ${k}^{(p)}(u,v)$ are differential operators of order $6-p$, written in $(u,v)$ coordinates \footnote{For $\om\neq 0$, there is a relation between $i_2$ in the type (\ref{i2-uv}) and 
the $y_6$ integral presented in \cite{TTW:2009}: $-\frac{27}{4}\,i_2 - h^3 = y_6$. 
This form of the integral $i_2$ leads to a minimal degrees of structure polynomials $P, Q$, see (11)-(12), 
of the polynomial algebra of integrals, unlike $y_6$.}
We must emphasize that the algebraic Hamiltonian $h_3$ and both the algebraic integrals
$i_1, i_2$ - all three operators - can be rewritten in terms of $g^{(3)}$-algebra generators.
Hence, this algebra is the hidden algebra of the $G_2/I_6$-rational, superintegrable model defined by
(\ref{HWol}) at $\om=0$.

For the sake of simplicity let us denote $h \equiv h_3$. 
It can be checked by direct calculations that
\[
 [h,i_1] = [h,i_2] = 0\ ,
\]
and the commutator of the integrals
\[
 [i_1,i_2] \equiv i_{12}\ ,
\]
does not vanish, evidently, it remains the integral,
\begin{equation*}
  [h,i_{12}]\ =\ 0\ .
\end{equation*}
In explicit form the integral $i_{12}$ is a rather complicated, a 7th order differential operator 
with polynomial coefficients which will be presented elsewhere.

As a result of tedious calculations, see \cite{vieyra2023wolfes} 
\footnote{In \cite{vieyra2023wolfes} the integral $i_{12}$ was mistakenly calculated as $[i_2,i_1]$ but 
was presented as $i_{12}=[i_1,i_2]$. This typo led to some differences in signs in the formulas 
presented in  \cite{vieyra2023wolfes} for the double commutators.}, the
double commutators $[i_1, i_{12}]$ and $[i_2, i_{12}]$ can be found as the 8th order
differential operator and the 12th order differential operator, respectively.
Surprisingly, they can be rewritten as very simple quartic polynomials in terms of
$(h, i_1, i_2, i_{12})$,
\begin{equation}
\label{G2-i1}
 [i_1, i_{12}]\ =\ \frac{32}{3}\,h^3\,i_1  + 144\,i_1\,i_2 - 72\,i_{12} +
 48\,(2\la+\nu-1)\,\bigg(2 (6\la+1)\,h^3  + 27\,(2\la+\nu+1)\,i_2 \bigg)\ ,
\end{equation}
and
\begin{equation}
\label{G2-i2}
 [i_2, i_{12}]\ =\ -\frac{32}{3}\,h^3\,i_2\ -\ 72\,i_2^2\ ,
\end{equation}
where the last commutator does not depend on $i_1$ and $i_{12}$, as well as parameters $\la,\nu$.
It is worth emphasizing that the dependence on the coupling constants (via the parameters $(\la, \nu)$) 
appears in the commutator $[{i}_1,{i}_{12}]$ {\it only}.
Hence, we arrive at {\it quartic} two-parametric polynomial algebra generated by the elements
$(h, {i}_1, {i}_2, {i}_{12})$. 
Syzygy 
\begin{align*}
 {i}_{12}^2 & =
   \frac{64}{3}\,(6\la+1))(2\la-1)\,h^6
   +\frac{64}{3}\,h^3\,i_1\,i_2
    + 192\,(12\,\la^2 + 2\,(3\nu-2)\,\la + (\nu - 9))\,h^3\,i_2
    \\ &
   -\frac{32}{3}\,h^3\,i_{12}
    - 144\,i_2\,i_{12} + 1296\,(2\la + \nu + 3)(2\la + \nu-3)\,i_2^2 + 144\, i_1\,i_2^2\ ,
\end{align*}
has the form of the sixth degree polynomial.
 At $\la=0$ this quartic polynomial algebra is reduced to the
{\it quartic} one-parametric polynomial algebra of integrals which corresponds
to the pure $A_2$ rational Calogero model (at $\om=0$).

\section{Tremblay-Turbiner-Winternitz (TTW) system (2009)}

\subsection{TTW Hamiltonian}
The Hamiltonian for the two-dimensional TTW system \cite{TTW:2009}, written in polar coordinates, is given by
\begin{equation}
\label{HTTW}
 H_k (r,\varphi;\om, \al, \beta)\ =\ -\pa_r^2 -
 \frac{1}{r}\pa_r - \frac{1}{r^2}\pa_{\varphi}^2  + \om^2 r^2 +
 \frac{\al k^2}{r^2 \cos^2 {k \varphi}} + \frac{\beta k^2}{r^2 \sin^2 {k \varphi}}\ ,
\end{equation}
where $k$, and $\al= a(a -1), \beta = b(b - 1) > - \frac{1}{4 k^2}$ and $\om > 0$ are parameters.
When $k$ in (\ref{HTTW}) takes integer values, the system is invariant wrt to dihedral group 
$I_{2k}$ transformations.

For integer $k$ the exact-solvability of (\ref{HTTW}) was demonstrated in \cite{TTW:2009}: energy eigenvalues 
appeared to be linear in quantum numbers. 
Superintegrability - the existence of two algebraically independent integrals - for $k=1,2,3,4$ 
was proved in \cite{TTW:2009} explicitly, 
for odd $k$ it was done by Quesne \cite{Quesne:2010}, while for integer $k$ 
(in fact, for rational $k$) by Kalnins-Kress-Miller \cite{Miller:2011}.

The Hamiltonian (\ref{HTTW}) to the best of the author's knowledge includes {\it all} published 
superintegrable and exactly-solvable planar systems in $E_2$ that allow 
a separation of variables in polar coordinates:
for $k=1$ this system coincides with the Smorodinsky-Winternitz system (SW-I), for $k=2$ 
it corresponds to the so-called $BC_2$ rational model, while for $k=3$ it describes 
the Wolfes model \cite{Wolfes:1974}, or, equivalently, the so-called $G_2/I_6$ rational 
model in the Hamiltonian reduction method nomenclature, 
in particular, if $\al=0$ this model is reduced to the $A_2$/3-body rational/Calogero 
model \cite{Calogero:1969}. 
For the general integer $k$ the Hamiltonian (\ref{HTTW}) corresponds to the 
$I_{2k}$ model in the Hamiltonian reduction method nomenclature \cite{OP:1983}.

The Hamiltonian (\ref{HTTW}) becomes the algebraic differential operator 
(with polynomial coefficients), which admits a Lie-algebraic interpretation. 
This operator can be achieved by making a gauge rotation,
\[
   h_k\ \equiv \ \Psi_0^{-1} (H_k - E_0) \Psi_0\ ,
\]
with the ground state eigenfunction,
\begin{equation}
\label{TTW-ground state}
    \Psi_0 \ =\ r^{(a+b)k}\ \cos^a {k \varphi}\ \sin^b {k \varphi}\ 
    e^{-\frac{\om r^2}{2}} \ ,
\end{equation}
and the ground state energy $E_0\ =\ 2\om [(a + b)k +1]$, with 
the subsequent change of variables,
\begin{equation}
\label{coord}
t\ =\ r^2\ ,\ u\ =\ r^{2k} \sin^2 {k\varphi}\ .
\end{equation}
These coordinates are the lowest order invariants of the dihedral group $I_{2k}$. 
The resulting gauge-transformed Hamiltonian takes the extremely simple 
algebraic form:
\begin{align}
  h_k\ &=   -4 t \pa^2_t - 8k u \pa^2_{tu} - 4k^2 t^{k-1} u \pa^2_u
\non \\ & 
\label{Lie}
    + 4[\om t - (a+b)k -1]\pa_t + [4\om k u - 2 k^2(2 b + 1)
    t^{k-1}]\pa_u \ .
\end{align}
This operator has infinitely-many finite-dimensional invariant subspaces 
${\cal P}_{\cal N}^{(s)}$ (\ref{space_r}) for $s \geq (k-1)$ and ${\cal N}=0,1,2,3\ldots $. 
For given $s$ these subspaces coincide with finite-dimensional representation spaces 
of the algebra $g^{(s)}$, they form the infinite flag ${\cal P}^{(s)}$, see (\ref{flag}). 

\subsection{Complete Integrability, first integral.}

It can be shown that the operator
\begin{equation}
\label{X_k}
  {\cal I}_1^{(k)} \equiv {\cal X}_k (\al, \beta)\ =\ -L_3^2 + 
  \frac{\al k^2}{\cos^2 {k \varphi}} + \frac{\beta k^2}{\sin^2 {k \varphi}}\ ,
\end{equation}
where $L_3 = \pa_{\varphi}$ is the 2D angular momentum, is an integral of motion for the TTW 
\hbox{system~\cite{Fris:1965, Wint:1966}}, see for discussion \cite{TTW:2009}.
Its existence is directly related to the separation of variables of the Schr\"odinger equation 
in polar coordinates. It can be said that the Hamiltonian (\ref{HTTW}) defines 
a completely-integrable system for any real $k \neq 0$. The gauge rotated 
\[
 x_k\ =\ \Psi_0^{-1} ({\cal X}_k - c_k) \Psi_0\ ,
\]
integral (\ref{X_k}), where $c_k=k^2(a + b)^2$ is the lowest eigenvalue of the integral 
${\cal X}_k$ (\ref{X_k}), takes the algebraic form
\begin{equation}
\label{X_k-alg}
    i_1^{(k)} \equiv x_k\ =\ -4\,k^2 u (t^k - u)\,\pa_u^2\ -\ 
    4\,k^2 [(b+\tfrac{1}{2})t^k - (a+b+1)u]\,\pa_u \ .
\end{equation}

\subsection{Superintegrability, second Integral}

In \cite{TTW:2009} the following conjecture was formulated:

{\bf
 Conjecture 1.
 \it A second integral of motion ${\cal Y}_{2k}$ of the order $2k$ (basic integral) 
 exists for the Hamiltonian (\ref{HTTW}) for all positive integer values of $k$. 
 In Cartesian coordinates ${\cal Y}_{2k}$ is a differential operator 
 of the order $2k$ with rational coefficients. The gauge transformation $\Psi_0^{-1} ({\cal Y}_k) \Psi_0$ 
 together with the change of variables (\ref{coord}) transforms ${\cal Y}_{2k}$ into the algebraic operator 
 ${y}_{2k}$ that has polynomial coefficients. The integral ${y}_{2k}$ is an element of the order $2k$ in 
 the enveloping algebra of the hidden algebra $g^{(k)}$. In particular, ${y}_{2k}$ contains the terms 
 $4^k [(J^1)^k-T_k](J^1)^k$ which fix $k=s$ in the hidden algebra $g^{(k)}$, see Section \ref{gk}.
 In the limit $\om = \al = 0$, the operator ${\cal Y}_{2k}(0,0,\beta)$ is reduced to the square 
 of an operator of order $k$.
}

All aspects of this Conjecture have been confirmed for $k=1,2,3,4$ for general $\om, \al, \beta$ 
as well as for $k=1,\ldots,6,8$ for $\om = \al = 0, \beta \neq 0$ in \cite{TTW:2009} 
\footnote{A few non-essential misprints for $k=2,3,4$ were found and then fixed later 
in \cite{Turbiner-SIGMA:2025}.}.

\subsection{Polynomial algebra of integrals of TTW model}

{\bf Conjecture 2} \cite{Turbiner-SIGMA:2025}}:

\noindent
{\it  The superintegrable and exactly solvable TTW system with integer index $k$ is characterized by 
the hidden algebra $g^{(k)}$ and 4-generated polynomial algebra of integrals of order $(k+1)$ with 
$H,{\cal I}_1,{\cal I}_2,{\cal I}_{12}$ as generating elements and with double commutators 
$[{\cal I}_1,{\cal I}_{12}]=P_{k+1}$,  $[{\cal I}_2,{\cal I}_{12}]=Q_{k+1}$, which can depend on ${\cal I}_{12}$ 
linearly; here $$Q_{k+1}= 8k^2\left((-)^k\, H^k {\cal I}_2\ -\ {\cal I}_2^2\right)+O(\om)\ .$$
The basic integrals ${\cal I}_1,{\cal I}_2$ have the form of the second order and $(2k)$-th order differential 
operators, respectively, as already conjectured in \cite{TTW:2009}.  
$H,{\cal I}_1,{\cal I}_2,{\cal I}_{12}$ are algebraically related: there exists syzygy
\[
  {\cal I}_{12}^2\ =\ {\cal R}_k(H,{\cal I}_1,{\cal I}_2,{\cal I}_{12})\ , 
\]
where ${\cal R}$ is $(4k+2)$-th order differential operator in dihedral group $I_{2k}$ invariants 
of the lowest order as variables. From another side, ${\cal R}_k$ is polynomial in $H$ of degree $2k$, where 
some terms can be predicted,
\[
   {\cal R}_k\ =\ 4\,k^4\,(2a+1)(2a-3)\,H^{2k}\ +\ (-)^{k+1} 8 k^2\,H^k\,(2{\cal I}_1\,{\cal I}_2 -
    {\cal I}_{12}) + 
\]
\[
    + 16 k^4 (-)^{k+1} (2 a^2 + 2ab - a + b - 9)\,H^k {\cal I}_2 +
\]   
\[   
   + 16 k^2 \Bigg({\cal I}_1\,{\cal I}_2^2 + k^2 \left((a+b)^2-9\right) {\cal I}_2^2 - {\cal I}_2\,{\cal I}_{12}\Bigg)\ +\ O(\om) \quad . 
\] 
}

This Conjecture has been confirmed for $k=1,2,3,4$ for general $\om, \al, \beta$ in \cite{Turbiner-SIGMA:2025}, 
as a result of extremely tedious calculations for $k=2,3,4$ by using MAPLE codes with subsequent check 
in Mathematica.

It seems evident that in general the obtained results will continue to hold for 
the classical TTW system \cite{TTW:2010} for \hbox{$k=1,2,3,4$} when the Lie brackets 
are replaced by Poisson brackets and a dequantization procedure is performed: 
$\pa \rar i p$, where $p$ is classical momentum.

\section{History: to the memory of Pavel Winternitz (AVT)}

When a monumental review paper by Miller-Post-Winternitz \cite{MPW:2013} was published,
Pavel Winternitz approached one of us (AVT) saying that after a critical analysis of this review 
\cite{MPW:2013} it becomes clear that (I) the formalism of quantum superintegrability should be 
further elaborated emphasizing the polynomial algebras of integrals and (II)  
the most important examples should be profoundly studied, in particular, the so-called 
Smorodinsky-Winternitz potentials I-II \cite{Fris:1965}. 
During 2013-2016 Pavel Winternitz and AVT worked out the formalism (I) and by 2018 
the Introduction and Generalities were mainly written in the present 
form~ 
\footnote{We preferred to make minimal modifications of these two parts 
as a memory to P Winternitz.} 
and we proceeded (slowly, without rush) to the Smorodinsky-Winternitz potentials I-II. 
After the sudden death of Pavel Winternitz in 2021 the review was continued 
with JC Lopez Vieyra and eventually completed by him together with AVT. 

\section*{Conclusions}

The goal of this short and grossly incomplete review is to make a detailed analysis 
of several basic two-dimensional quantum superintegrable systems in flat space 
in order to clarify and extend the Montreal conjecture \cite{TTW:2001} about 
the exact solvability of superintegrable systems.
The Hamiltonian of each superintegrable system is taken as a superposition of 
the Laplace-Beltrami operator with the flat space metric and a scalar potential.

We showed that the Smorodinsky-Winternitz potentials Cases I-II, the Fokas-Lagerstrom model, 
the 3-body singular rational Calogero and Wolfes models, the TTW system at integer index $k$, 
all of them, additionally to exact-solvability, have polynomial eigenfunctions in the invariants 
of their symmetry group 
taken as variables, they admit an algebraic form of differential operators with polynomial coefficients, 
they possess a hidden algebraic structure, and each of them has a polynomial, 4-generated infinite-dimensional 
algebra of integrals. Being realized in differential operators the polynomial algebra is 
``constrainted" due to the existence of syzygy - four generating elements are algebraically related. 
Due to the high level of technicality we were forced to recalculate many results known in the literature in order to fix errors, typos, misprints and unify notations.

It must be noted that the existence of the algebraic form of superintegrable system
allows us to construct the Fock space representation in the operators/letters 
$a_{1,2}, b_{1,2}$, which are elements of five-dimensional Heisenberg algebra $h_5$. 
These operators can be realized as finite-difference and/or shift operators on uniform 
or exponential lattice. Thus, we arrive at the isospectral systems on uniform, 
exponential lattices and mixed, uniform-exponential lattices, see e.g. \cite{TLG:2024} 
and references therein.

The analysis of three quantum superintegrable systems in non-flat space 
(the Higgs I-II potentials, thus, the spherical/hyperbolic harmonic oscillator 
and the Kepler-Coulomb problem \cite{Higgs:1979}, the reduced Coulomb problem 
\cite{TER:2023}) supports the validity of the original Montreal conjecture - 
they are exactly-solvable with explicitly known energy eigenvalues, 
they also possess the polynomial algebras of integrals \cite{Bonatsos:1993}, 
\cite{TER:2023}. Recently, a new generalization of the TTW system into the spaces 
of constant curvature $S_2$ and $H_2$ was proposed \cite{delOlmo:2025}: 
it remained superintegrable 
(both integrals might be constructed explicitly at least for some index $k$) 
and exactly-solvable, while the question about polynomiality 
of the algebra of integrals remained open.

\section*{Acknowledgments}

AVT is thankful to CRM, University of Montreal for its kind hospitality extended to him 
where this work was started and progressed in 2013-6 together with P Winternitz, 
who passed away on Febr.13, 2021, and it never was completed.
The authors discussed the subject with W Miller Jr (University of Minnesota) 
on multiple occasions, who passed away on Oct.29, 2023. 
This paper is dedicated to the memory of Pavel Winternitz and Willard Miller Jr, 
two giants of mathematical physics. Both present authors feel strongly their absence. 

AVT expresses the gratitude to G Sterman, R Schrock, B McCoy (YITP, Stony Brook) 
for their interest to the subject and kind invitation to present a regular seminar 
at YITP on this subject, where the idea to write a review about algebras of integrals 
was born - it is a part of the present review. 
The authors thank Dr K K Phua (World Scientific) for the kind invitation to write this review paper. 
We are thankful to A M Escobar-Ruiz for participation in the early stage of work and numerous discussions. 
JCLV was supported in part by PASPA project for his sabbatical stay in Southern Methodist University (Dallas, USA), where significant part of this review was prepared.

This work is supported in part by the PAPIIT grants {\bf IN108815, IN113022, IN104125} and 
CONACyT grants {\bf 166189, A1-S-17364}~(Mexico).

\end{document}